\documentclass[10pt,fleqn]{article}
\usepackage[utf8]{inputenc}
\usepackage[T1]{fontenc}
\usepackage[top=1in,bottom=1in,left=1.2in,right=1.2in]{geometry}
\usepackage{authblk,hyperref,xcolor,units}
\usepackage{amsmath,amsfonts,amssymb}

\numberwithin{equation}{section}

\newcommand{\ket}[1]{\rvert#1\rangle}
\newcommand{\bra}[1]{\langle #1\rvert}
\newcommand{\braket}[2]{\langle #1\rvert#2\rangle}
\newcommand{\Braket}[3]{\bra{#1}#2\ket{#3}}

\newcommand{\osp}{\mathfrak{osp}}
\newcommand{\sll}{\mathfrak{sl}}
\newcommand{\su}{\mathfrak{su}}

\newcommand{\BLUE}[1]{{\color{blue} #1}}
\renewcommand{\BLUE}[1]{{\color{black} #1}}                     
\newcommand{\eu}{\BLUE{\epsilon_1}}
\newcommand{\ed}{\BLUE{\epsilon_2}}
\newcommand{\eud}{\BLUE{\epsilon_{12}}}
\newcommand{\cN}{\mathcal{N}}
\newcommand{\cK}{\mathcal{K}}

\newcommand*\pFqskip{8mu}
\catcode`,\active
\newcommand*\pFq{\begingroup
        \catcode`\,\active
        \def ,{\mskip\pFqskip\relax}%
        \dopFq}
\catcode`\,12
\def\dopFq#1#2#3#4#5{%
        {}_{#1}F_{#2}\biggl(\genfrac..{0pt}{}{#3}{#4};#5\biggr)%
        \endgroup}

\makeatletter
\let\cite\relax
\DeclareRobustCommand{\cite}{%
  \let\new@cite@pre\@gobble
  \@ifnextchar[\new@cite{\@citex[]}}
\def\new@cite[#1]{\@ifnextchar[{\new@citea{#1}}{\@citex[#1]}}
\def\new@citea#1{\def\new@cite@pre{#1}\@citex}
\def\@cite#1#2{[{\new@cite@pre\space#1\if\relax\detokenize{#2}\relax\else, #2\fi}]}
\makeatother

\title{Superintegrability and the dual $-1$ Hahn algebra in superconformal quantum mechanics}

\author[1]{Pierre-Antoine Bernard}
\author[1,2]{Julien Gaboriaud\footnote{\url{julien.gaboriaud@umontreal.ca}}}
\author[1,2]{Luc Vinet}

\affil[1]{D\'epartement de Physique, Universit\'e de Montr\'eal, Montr\'eal (Qu\'ebec), H3C 3J7, Canada}
\affil[2]{Centre de Recherches Math\'ematiques, Universit\'e de Montr\'eal, P.O. Box 6128, Centre-ville Station, Montr\'eal, Canada, H3C 3J7}
\date{\today}

\begin{document}

\maketitle
\thispagestyle{empty}
\hrule
\begin{abstract}\noindent
A two-dimensional superintegrable system of singular oscillators with internal degrees of freedom is identified and exactly solved. Its symmetry algebra is seen to be the dual $-1$ Hahn algebra which describes the bispectral properties of the polynomials with the same name that are essentially the Clebsch-Gordan coefficients of the superconformal algebra $\mathfrak{osp}(1|2)$. It is also shown how this superintegrable model is obtained under dimensional reduction from a set of uncoupled harmonic oscillators in four dimensions.
\end{abstract}
\hrule

\section{Introduction}
This paper introduces a simple superintegrable model in two-dimensions with internal degrees of freedom. Its Hamiltonian defined on $L^{2}(\mathbb{R}^{2},\mathbb{C}^{4})$ belongs to a realization of the superconformal algebra $\sll(2|1)$ and reads
\begin{align}\label{eq:H}
 H=-\frac{1}{2}\left(\frac{\partial^{2}}{\partial\rho_1^{2}}+\frac{\partial^{2}}{\partial\rho_2^{2}}\right)+\frac{1}{2}\left(\rho_1^{2}+\rho_2^{2}\right)+\frac{k_1(k_1-\sigma_3\otimes1)}{2\rho_1^{2}}+\frac{k_2(k_2-1\otimes\sigma_3)}{2\rho_2^{2}},
\end{align}
with $\sigma_3=\left(\begin{smallmatrix} 1~ & 0 \\[.25em] 0~ & -1 \end{smallmatrix}\right)$ the standard Pauli matrix. The symmetry algebra generated by the constants of motion includes the not-so-familiar dual $-1$ Hahn algebra \cite{Genest2013}.

Superintegrable models in $d$ dimensions have the property of admitting more than $d$ (independent) constants of motion which hence form a non-Abelian symmetry algebra. In the scalar case, they are called maximally superintegrable if the number of these constants is equal to $2d-1$ when the Hamiltonian is included. They are typically exactly solvable. Interest in these systems is high. Some of the best known examples like the harmonic oscillator or the Kepler/Coulomb problem are discussed in textbooks and are of central use in Physics. Chiefly, these models are laboratories to study diverse expressions of extended symmetries. Therein lies the motivation for their systematic exploration and identification. The reader may consult \cite{Miller2013} for a recent review.

One constructive approach to obtain superintegrable models that we shall use here combines one-dimensional systems equipped with ladder operators \cite{Letourneau1995,Marquette2010}. Consider for concreteness the singular harmonic oscillator $\tilde{H}=p^{2}/(2m)+\omega \rho^{2}+\lambda\rho^{-2}$ which can be cast as a generator in a realization of $\su(1,1)$ and which hence exhibits conformal symmetry \cite{DeAlfaro1976}. From the properties of the discrete series representations of $\su(1,1)$, the spectrum of $H$ is known to be equidistantly spaced and therefore linear in a quantum number $n$. Adding two such realizations say in the variable $\rho_1$ and $\rho_2$ to form a two-dimensional system leads to a Hamiltonian with manifest degeneracies. The corresponding constants of motion are readily constructed as products of the $\su(1,1)$ raising and lowering operators from each of the two summands that leave the total energy unchanged. These constants are seen to generate \cite{Letourneau1995,Genest2014} the Higgs algebra \cite{Higgs1979,Zhedanov1992,Bonatsos1995} which is isomorphic to the dual Hahn algebra \cite{Granovskii1991,Granovskii1988,Frappat2019a}. The latter is an example of the quadratic algebras \cite{Zhedanov1991} that are associated to the hypergeometric orthogonal polynomials of the Askey scheme \cite{Koekoek2010}. These algebras are realized by the bispectral operators of the corresponding families of polynomials. In the case of the dual Hahn algebra, the associated dual Hahn polynomials are essentially the Clebsch-Gordan coefficients of $\su(1,1)$.

We wish to examine if this approach extends fruitfully to the supersymmetric context. Can one combine one-dimensional superconformal Hamiltonian \cite{Fubini1984} with $\osp(1|2)$ as dynamical algebra to obtain a superintegrable model? Positive indications arise from the fact that the representations of $\osp(1|2)$ that belong to the discrete series imply a linear spectrum for the Cartan generator which leads, here also, to degenerate situations for this operator in combined representations. We shall call upon operators acting on vector-valued functions to provide the appropriate realizations. Interestingly, we shall thus find a model with internal degrees of freedom that has for symmetry algebra the one associated to the dual $-1$ Hahn polynomials. These polynomials have been singled out \cite{Tsujimoto2013} as the $q\to-1$ limit of the dual $q$-Hahn polynomials \cite{Koekoek2010} and shown to be basically the Clebsch-Gordan coefficients of $\osp(1|2)$ \cite{Genest2013}.

The present study has similarities with the analysis of the Dunkl oscillator in the plane that also brings on supersymmetry \cite{Genest2012,Genest2014a,Genest2013a}. In this case the relevant realizations of $\osp(1|2)$ are the parabosonic \cite{Rosenblum1994,Mukunda1980} ones constructed in terms of Dunkl and reflection operators \cite{Dunkl1991}. A deformation of $\su(2)$ obtained by extending the Schwinger construction to Dunkl creation and annihilation operators was identified as the symmetry algebra responsible for the superintegrability of the Dunkl oscillator in two dimensions. It was called the Schwinger--Dunkl algebra $sd(2)$ but it actually coincides with the dual $-1$ Hahn algebra.

The upshot of the present paper is that there is a simple superintegrable model with internal degrees of freedom that possesses the same dual $-1$ Hahn symmetry algebra and the attractive feature of not involving reflection operators.

A further observation is that this new model can be obtained as well through dimensional reduction. It was shown in \cite{Rodriguez2008} that harmonic oscillators in $2d$ dimensions can be reduced to maximally superintegrable systems of singular oscillators in $d$ dimensions with integrals of motion inherited from those in the higher dimensions. This will be seen to prevail for the model with dual $-1$ Hahn symmetry upon projecting uncoupled harmonic oscillators with internal degrees of freedom from four to two dimensions.

The paper is structured as follows. Section \ref{sec:osp12} recalls facts about $\osp(1|2)$ and its representations of the discrete series. Section \ref{sec:model} introduces the relevant realizations of $\osp(1|2)$ in terms of matrix differential operators and the two-dimensional superconformal Hamiltonian of interest. This Hamiltonian is shown to be superintegrable in Section \ref{sec:symalgebra} and its symmetry algebra is identified as the dual $-1$ Hahn algebra. The wavefunctions separated in both Cartesian and polar coordinates are presented in Section \ref{sec:solutions}. The total $\osp(1|2)$ Casimir element, as one of the constants of motion, is also associated to separation in polar coordinates. Its eigenfunctions are obtained in Section \ref{sec:overlaps} and their overlaps with the wavefunctions separated in Cartesian coordinates are shown to be given in terms of the dual $-1$ Hahn polynomials. Finally, how the superintegrable singular oscillator with internal degrees of freedom can be derived via dimensional reduction is the subject of Section \ref{sec:reduction}. Concluding remarks are found in Section \ref{sec:conclusion}. Appendix \ref{sec:dualHahn} gathers the main properties of the dual $-1$ Hahn polynomials and Appendix \ref{sec:differential} collects technical details on the solutions of relevant differential equations.

\section{A review of $\osp(1|2)$}\label{sec:osp12}
The Lie superalgebra $\osp(1|2)$ can be presented as the algebra with generators $A_0$, $A_\pm$ and an involution $P$ encoding the $\mathbb{Z}_2$-grading of the superalgebra ($P$ commutes with the even elements and anticommutes with the odd elements). These obey the relations
\begin{align}
 \{A_+,A_-\}=2A_0,\qquad [A_0,A_\pm]=\pm A_\pm,\qquad  [P,A_0]=0,\qquad \{P,A_\pm\}=0.
\end{align}
The sCasimir of $\osp(1|2)$
\begin{align}
 S=A_+A_--A_0+\tfrac{1}{2}
\end{align}
belongs to the universal enveloping algebra of $\osp(1|2)$ and satisfies the following relation
\begin{align}
 [S,A_0]=\{S,A_{\pm}\}=0.
\end{align}
In other words, the sCasimir commutes with even generators and anticommutes with odd generators. One can then form a Casimir element by multiplying $S$ with $P$:
\begin{align}
 Q=\left(A_+A_--A_0+\tfrac{1}{2}\right)P=\tfrac{1}{2}\left([A_+,A_-]+1\right)P.
\end{align}
Positive infinite-dimensional discrete series representations are labelled by $(\mu,\epsilon)$, with $\mu\geq0$, $\epsilon=\pm1$ and the actions of the generators on the associated basis vectors are given by
\begin{align}
 A_0\ket{\mu,n,\epsilon}&=(n+\mu+\tfrac{1}{2})\ket{\mu,n,\epsilon},\label{eq:osp12a0}\\
 A_+\ket{\mu,n,\epsilon}&=\sqrt{[n+1]_\mu}\ket{\mu,n+1,\epsilon},\label{eq:osp12ap}\\
 A_-\ket{\mu,n,\epsilon}&=\sqrt{[n]_\mu}\ket{\mu,n-1,\epsilon},\label{eq:osp12am}\\
 P\ket{\mu,n,\epsilon}&=\epsilon(-1)^{n}\ket{\mu,n,\epsilon},\label{eq:actionP}
\end{align}
with the \textit{mu}-numbers $[n]_\mu$ defined as the following
\begin{align}
 [n]_\mu=n+\mu\left(1-(-1)^{n}\right).
\end{align}
The Casimir element acts as a multiple of the identity on these irreps
\begin{align}
 Q\ket{\mu,n,\epsilon}=-\epsilon\mu\ket{\mu,n,\epsilon}.
\end{align}
In the realizations of $\osp(1|2)$ that we shall soon provide, $A_0$ will be interpreted as the Hamiltonian.

\subsection{The Clebsch-Gordan problem of $\osp(1|2)$}
We now turn to the study of the recoupling problem of two copies of $\osp(1|2)$. We first introduce the coproduct $\Delta:\osp(1|2)\to\osp(1|2)\otimes\osp(1|2)$, a coassociative algebra morphism that acts as follows on the generators:
\begin{align}
\begin{aligned}
 \Delta(A_0)&=A_0\otimes1+1\otimes A_0\\
 \Delta(A_\pm)&=A_\pm\otimes P+1\otimes A_\pm\\
 \Delta(P)&=P\otimes P
\end{aligned}\
\begin{aligned}
 &=A_0^{(1)}+A_0^{(2)},\\
 &=A_\pm^{(1)}P^{(2)}+A_\pm^{(2)},\\
 &=P^{(1)}P^{(2)}.
\end{aligned}
\end{align}
As a result
\begin{align}\label{eq:Q12algebra}
 \Delta(Q)=Q^{(12)}=(A_-^{(1)}A_+^{(2)}-A_+^{(1)}A_-^{(2)})P^{(1)}+Q^{(1)}P^{(2)}+Q^{(2)}P^{(1)}-\tfrac{1}{2}P^{(1)}P^{(2)}.
\end{align}
Let us now look at the recoupling of two irreducible representations of $\mathfrak{osp}(1|2)$ denoted $(\mu_1,\epsilon_1)$ and $(\mu_2,\epsilon_2)$, following \cite{Bergeron2016}. There are two natural bases to consider.

To the direct product representation $(\mu_1,\epsilon_1)\otimes(\mu_2,\epsilon_2)$ are associated the basis vectors given by $\ket{\mu_1,n_1,\epsilon_1}\otimes\ket{\mu_2,n_2,\epsilon_2}$ that diagonalize the Casimir elements $Q^{(1)}$ and $Q^{(2)}$, $A_0\otimes1$, $1\otimes A_0$ and $P\otimes1$, $1\otimes P$. It should be noted that it is equivalent to diagonalize $A_0\otimes1$, $\Delta(A_0)$, $P\otimes1$, $\Delta(P)$ instead of the previous four (this will be used later when we consider the Clebsch-Gordan algebra associated to $\osp(1|2)$). All these elements act as follows on the basis vectors
\begin{align}
\begin{aligned}
 (Q\otimes1)&\ket{\mu_1,n_1,\epsilon_1}\otimes\ket{\mu_2,n_2,\epsilon_2}=\\
 (1\otimes Q)&\ket{\mu_1,n_1,\epsilon_1}\otimes\ket{\mu_2,n_2,\epsilon_2}=\\
 (A_0\otimes1)&\ket{\mu_1,n_1,\epsilon_1}\otimes\ket{\mu_2,n_2,\epsilon_2}=\\
 (1\otimes A_0)&\ket{\mu_1,n_1,\epsilon_1}\otimes\ket{\mu_2,n_2,\epsilon_2}=\\
 \Delta(A_0)&\ket{\mu_1,n_1,\epsilon_1}\otimes\ket{\mu_2,n_2,\epsilon_2}=\\
 (P\otimes1)&\ket{\mu_1,n_1,\epsilon_1}\otimes\ket{\mu_2,n_2,\epsilon_2}=\\
 (1\otimes P)&\ket{\mu_1,n_1,\epsilon_1}\otimes\ket{\mu_2,n_2,\epsilon_2}=\\
 \Delta(P)&\ket{\mu_1,n_1,\epsilon_1}\otimes\ket{\mu_2,n_2,\epsilon_2}=
\end{aligned}\
\begin{aligned}
 &-\epsilon_1\mu_1\ket{\mu_1,n_1,\epsilon_1}\otimes\ket{\mu_2,n_2,\epsilon_2},\\
 &-\epsilon_2\mu_2\ket{\mu_1,n_1,\epsilon_1}\otimes\ket{\mu_2,n_2,\epsilon_2},\\
 &(n_1+\mu_1+\tfrac{1}{2})\ket{\mu_1,n_1,\epsilon_1}\otimes\ket{\mu_2,n_2,\epsilon_2},\\
 &(n_2+\mu_2+\tfrac{1}{2})\ket{\mu_1,n_1,\epsilon_1}\otimes\ket{\mu_2,n_2,\epsilon_2},\\
 &(n_1+n_2+\mu_1+\mu_2+1)\ket{\mu_1,n_1,\epsilon_1}\otimes\ket{\mu_2,n_2,\epsilon_2},\\
 &\epsilon_1(-1)^{n_1}\ket{\mu_1,n_1,\epsilon_1}\otimes\ket{\mu_2,n_2,\epsilon_2},\\
 &\epsilon_2(-1)^{n_2}\ket{\mu_1,n_1,\epsilon_1}\otimes\ket{\mu_2,n_2,\epsilon_2},\\
 &\epsilon_1\epsilon_2(-1)^{n_1+n_2}\ket{\mu_1,n_1,\epsilon_1}\otimes\ket{\mu_2,n_2,\epsilon_2}.
\end{aligned}
\end{align}
To the irreducible components $(\mu_{12},\epsilon_{12})$ of the decomposition of the direct product representation are attached the basis vectors $\ket{\mu_{12},n_{12},\epsilon_{12}}$. The diagonal operators are $\Delta(Q)$, $\Delta(A_0)$, $\Delta(P)$:
\begin{align}
\begin{aligned}
 \Delta(Q)&\ket{\mu_{12},n_{12},\epsilon_{12}}=\\
 \Delta(A_0)&\ket{\mu_{12},n_{12},\epsilon_{12}}=\\
 \Delta(P)&\ket{\mu_{12},n_{12},\epsilon_{12}}=
\end{aligned}\
\begin{aligned}
 &-\epsilon_{12}\mu_{12}\ket{\mu_{12},n_{12},\epsilon_{12}},\\
 &(n_{12}+\mu_{12}+\tfrac{1}{2})\ket{\mu_{12},n_{12},\epsilon_{12}},\\
 &\epsilon_{12}(-1)^{n_{12}}\ket{\mu_{12},n_{12},\epsilon_{12}}.
\end{aligned}
\end{align}
Owing to the decomposition \cite{Genest2015a}
\begin{align}
 (\mu_1,\epsilon_1)\otimes(\mu_2,\epsilon_2)=\bigoplus_{j=0}^{\infty}(\mu_1+\mu_2+j+\tfrac{1}{2},(-1)^{j}\epsilon_1\epsilon_2)
\end{align}
one directly obtains that
\begin{align}
 \mu_{12}=\mu_1+\mu_2+j+\tfrac{1}{2},\qquad \epsilon_{12}=(-1)^{j}\epsilon_1\epsilon_2,\qquad j\in\mathbb{N}.
\end{align}
The Clebsch-Gordan coefficients are then defined as the expansion coefficients between these two bases
\begin{align}
 \ket{\mu_{12},n_{12},\epsilon_{12}}=\sum_{n_1,n_2}\mathcal{C}_{n_{12},j}^{n_1,n_2}\ket{\mu_1,n_1,\epsilon_1}\otimes\ket{\mu_2,n_2,\epsilon_2}.
\end{align}
It can be shown \cite{Genest2013,Tsujimoto2011} that the Clebsch-Gordan coefficients are expressible in terms of the dual $-1$ Hahn polynomials. These Clebsch-Gordan coefficients will appear in Section \ref{sec:overlaps} as overlaps of the solutions of our spinorial superintegrable system separated in both the Cartesian and polar coordinates.

\section{A spinorial realization of $\osp(1|2)$}\label{sec:model}
Dunkl realizations of $\osp(1|2)$ have been studied extensively \cite{Mukunda1980,Rosenblum1994,Genest2012,Genest2014a}. These realizations are built with reflection operators and were taken to act on scalar wavefunctions. In this paper we shall focus instead on a realization with spin (internal) degrees of freedom which offers a valuable alternative perspective. The wavefunctions of the system are then given in terms of spinors. In the remainder of this paper, we will refer to this realization as the spinorial realization of $\osp(1|2)$.

\subsection{The spinorial model}
Recall the usual Pauli matrices $\{\sigma_1,\sigma_2,\sigma_3\}$ and form
\begin{align}\label{eq:ApmA0}
 A_{\pm}=\frac{1}{\sqrt{2}}\left[\sigma_1\left(\rho\mp\frac{\partial}{\partial\rho}\right)\mp i \sigma_2\frac{k}{\rho}\right],\qquad A_0=-\frac{1}{2}\frac{d^{2}}{d\rho^{2}}+\frac{\rho^{2}}{2}+\frac{k(k-\sigma_3)}{2\rho^{2}}.
\end{align}
Using $\sigma_a\sigma_b=\delta_{ab}+i\epsilon_{abc}\sigma_c$, one easily checks that
\begin{align}
 \{A_+,A_-\}=2A_0,\qquad [A_0,A_{\pm}]=\pm A_\pm.
\end{align}
We shall interpret $H=A_0$ as the Hamiltonian of this spinorial model.
Note that parity $P:\rho\mapsto-\rho$ satisfies the remaining relations of $\osp(1|2)$
\begin{align}\label{eq:relationsP}
 [P,A_0]=\{P,A_{\pm}\}=0.
\end{align}
Using the above, it is also seen that the Casimir element in this realization takes the form
\begin{align}
 Q=-k\sigma_3P.
\end{align}
Recalling that $Q$ has eigenvalue $-\epsilon\mu$ in a discrete series irrep of $\osp(1|2)$ and taking $\mu=k$, we hence take
\begin{align}\label{eq:Psigma3}
 P\sigma_3=\sigma_3 P=\epsilon
\end{align}
and it is seen that the identification $P=\epsilon\sigma_3$ in irreps is consistent with the relations \eqref{eq:relationsP}.

What does this tell us about the wavefunctions? Recall that $\sigma_3$ has eigenvalue $+1$ for spin up ($\uparrow$) and $-1$ for spin down ($\downarrow$). On the other hand, $P$ encodes the parity of the wavefunction: it has eigenvalue $+1$ for even wavefunctions and $-1$ for odd wavefunctions. The eigenvalues then combine like this:
\begin{align}\begin{aligned}\label{eq:conditionsk}
 \text{even wavefunction}\quad\&\quad\text{spin $\uparrow$}\qquad&\longrightarrow\qquad\text{eigenvalue of $\sigma_3 P=+1=\epsilon$},\\
 \text{odd wavefunction}\quad\&\quad\text{spin $\downarrow$}\qquad&\longrightarrow\qquad\text{eigenvalue of $\sigma_3 P=+1=\epsilon$},\\
 \text{even wavefunction}\quad\&\quad\text{spin $\downarrow$}\qquad&\longrightarrow\qquad\text{eigenvalue of $\sigma_3 P=-1=\epsilon$},\\
 \text{odd wavefunction}\quad\&\quad\text{spin $\uparrow$}\qquad&\longrightarrow\qquad\text{eigenvalue of $\sigma_3 P=-1=\epsilon$}.
\end{aligned}\end{align}
In other words, for our spinorial model and for a given $k$, the set of scenarios \eqref{eq:conditionsk} provides us with an interpretation of the sign $\epsilon$ labelling the representations. Either (a) $\epsilon=+1$ and then there is a matching between the even wavefunctions and the spin up, and between the odd wavefunctions and the spin down, or (b) $\epsilon=-1$ and then there is a matching between the odd wavefunctions and the spin up, and between the even wavefunctions and the spin down.

This interpretation is in keeping with the fact that once we identify $P$ with $\epsilon\sigma_3$ in irreps, we have from \eqref{eq:actionP} that
\begin{align}\label{eq:contradiction}
 P\ket{\mu,n,\epsilon}=\epsilon(-1)^{n}\ket{\mu,n,\epsilon}=\epsilon\sigma_3\ket{\mu,n,\epsilon},\qquad\text{and hence}\qquad\sigma_3\ket{\mu,n,\epsilon}=(-1)^{n}\ket{\mu,n,\epsilon}.
\end{align}
This connects the spin with the energies which are known to depend on parity.

\subsection{The $\osp(2|2)$ dynamical algebra}
The dynamical algebra of this spinorial model further realizes $\osp(2|2)\simeq\mathfrak{sl}(2|1)$. This superalgebra can be presented \cite{Frappat1996} in terms of four bosonic generators $E^{\pm}$, $\bar H$, $Z$ and two pairs of fermionic generators $F^{\pm}$, $\bar{F}^{\pm}$ obeying the relations
\begin{align}\label{eq:sl21}
\begin{aligned}{}
 [\bar H,E^{\pm}]&=\pm E^{\pm},\\
 [E^{+},E^{-}]&=2\bar H,\\[0.5em]
 [Z,E^{\pm}]&=0=[Z,\bar H],\\[1.em]
 \{F^{\pm},\bar{F}^{\mp}\}&=Z\mp\bar H,\\
 \{F^{\pm},\bar{F}^{\pm}\}&=E^{\pm},\\[.5em]
 \{F^{\pm},F^{\pm}\}&=0=\{\bar{F}^{\pm},\bar{F}^{\pm}\},\\
 \{F^{\pm},F^{\mp}\}&=0=\{\bar{F}^{\pm},\bar{F}^{\mp}\},
\end{aligned}\hspace{6em}
\begin{aligned}{}
 [\bar H,F^{\pm}]&=\pm\frac{1}{2}F^{\pm},\\
 [Z,F^{\pm}]&=\phantom{-}\frac{1}{2}F^{\pm},\\
 [\bar H,\bar{F}^{\pm}]&=\pm\frac{1}{2}\bar{F}^{\pm},\\
 [Z,\bar{F}^{\pm}]&=-\frac{1}{2}\bar{F}^{\pm},\\[.5em]
 [E^{\pm},F^{\mp}]&=-F^{\pm},\\
 [E^{\pm},\bar{F}^{\mp}]&=\phantom{-}\bar{F}^{\pm},\\
 [E^{\pm},F^{\pm}]&=0=[E^{\pm},\bar{F}^{\pm}].\\
\end{aligned}
\end{align}
Here is how the algebra is realized in our model. As previously defined in \eqref{eq:ApmA0}, the element $A_0$ is a bosonic generator and $A_{\pm}$ are fermionic generators.
Introduce $Y$, which commutes with $A_0$ and is realized by
\begin{align}
 Y=\frac{\sigma_3}{2i}.
\end{align}
This element leads to another $\osp(1|2)$ realization. Indeed, we can form supercharges tilde as follows
\begin{align}
 \tilde{A}_{\pm}=[A_{\pm},Y]=\frac{1}{\sqrt{2}}\left[-\sigma_2\left(\rho\mp\frac{\partial}{\partial\rho}\right)\mp i \sigma_1\frac{k}{\rho}\right].
\end{align}
Those supercharges also obey the $\osp(1|2)$ defining relations:
\begin{align}
 \{\tilde{A}_+,\tilde{A}_-\}=2A_0,\qquad [A_0,\tilde{A}_{\pm}]=\pm \tilde{A}_\pm,\qquad [P,A_0]=\{P,\tilde{A}_{\pm}\}=0.
\end{align}
The dynamical algebra hence contains the two $\osp(1|2)$ subalgebras mentioned above and can be identified as $\mathfrak{sl}(2|1)$ by mapping its generators in the following way:
\begin{align}
 \bar H&=\frac{1}{2}A_0=\frac{1}{4}\left(-\frac{\partial^{2}}{\partial\rho^{2}}+\frac{k(k-\sigma_3)}{\rho^{2}}+\rho^{2}\right),\label{eq:osp22H}\\
 Z&=-\frac{1}{2}\left(k+iY\right)=-\frac{1}{2}\left(k+\frac{\sigma_3}{2}\right),\\
 F^{+}&=\frac{\phantom{-i}(\tilde{A_+}-iA_+)}{2\sqrt{2}}=\phantom{-}\frac{i}{4}(\sigma_1-i\sigma_2)\left(\frac{\partial}{\partial\rho}-\frac{k}{\rho}-\rho\right),\\
 F^{-}&=\frac{-i(\tilde{A_-}-iA_-)}{2\sqrt{2}}=-\frac{1}{4}(\sigma_1-i\sigma_2)\left(\frac{\partial}{\partial\rho}-\frac{k}{\rho}+\rho\right),\\
 \bar{F}^{+}&=\frac{\phantom{-}i(\tilde{A_+}+iA_+)}{2\sqrt{2}}=\phantom{-}\frac{1}{4}(\sigma_1+i\sigma_2)\left(\frac{\partial}{\partial\rho}+\frac{k}{\rho}-\rho\right),\\
 \bar{F}^{-}&=\frac{-\phantom{i}(\tilde{A_-}+iA_-)}{2\sqrt{2}}=-\frac{i}{4}(\sigma_1+i\sigma_2)\left(\frac{\partial}{\partial\rho}+\frac{k}{\rho}+\rho\right),\\
 E^{\pm}&=\{F^{\pm},\bar{F}^{\pm}\}=-\frac{i}{4}\left[-\frac{\partial^{2}}{\partial\rho^{2}}+\frac{k(k-\sigma_3)}{\rho^{2}}-\rho^{2}\pm\left(1+2\rho\frac{\partial}{\partial\rho}\right)\right].\label{eq:osp22Epm}
\end{align}
The defining relations \eqref{eq:sl21} are checked directly.

In the following we shall focus only on the $\osp(1|2)$ realization given in \eqref{eq:ApmA0}.

\section{Constants of motion and the dual $-1$ Hahn algebra}\label{sec:symalgebra}
In this Section we will show that the dual $-1$ Hahn algebra is the symmetry algebra that accounts for the superintegrability of the two-dimensional model with internal degrees of freedom \eqref{eq:H}.

\subsection{Two-dimensional model}
Using the $\osp(1|2)$ coproduct, we can form the following two-dimensional realization:
\begin{align}
A_\pm^{(12)}&=A_\pm\otimes P+1\otimes A_\pm\nonumber\\
             &=\ed\frac{1}{\sqrt{2}}\left[(\sigma_1\otimes\sigma_3)\left(\rho_1\mp\frac{d}{d\rho_1}\right)\mp i(\sigma_2\otimes\sigma_3)\frac{k_1}{\rho_1}\right]\\
             &\phantom{\epsilon_2}+\frac{1}{\sqrt{2}}\left[\phantom{\sigma}(1\otimes\sigma_1)\left(\rho_2\mp\frac{d}{d\rho_2}\right)\mp i\phantom{\sigma}(1\otimes\sigma_2)\frac{k_2}{\rho_2}\right]\nonumber\\[.5em]
 A_0^{(12)}&=-\frac{1}{2}\left(\frac{d^{2}}{d\rho_1^{2}}+\frac{d^{2}}{d\rho_2^{2}}\right)+\frac{1}{2}(\rho_1^{2}+\rho_2^{2})+\frac{k_1(k_1-\sigma_3\otimes1)}{2\rho_1^{2}}+\frac{k_2(k_2-1\otimes\sigma_3)}{2\rho_2^{2}}.
\end{align}
The involution $P$ is mapped to
\begin{align}
 P^{(12)}=P^{(1)}P^{(2)}=\eu\ed(\sigma_3\otimes\sigma_3).
\end{align}
These elements still satisfy the $\osp(1|2)$ algebra relations since they were obtained from the coproduct morphism.

Now introduce the gamma matrices
\begin{align}
 \gamma_1&=\sigma_1\otimes\sigma_3,\qquad
 \gamma_2=\sigma_2\otimes\sigma_3,\qquad
 \gamma_3=1\otimes\sigma_1,\qquad
 \gamma_4=1\otimes\sigma_2.
\end{align}
These obey the Clifford algebra relations
\begin{align}
 \{\gamma_a,\gamma_b\}=2\delta_{ab}.
\end{align}
Denoting $\Sigma_{ab}=i\gamma_a\gamma_b$, we form the two spin operators
\begin{align}
 \Sigma_{12}=-(\sigma_3\otimes1),\qquad \Sigma_{34}=-(1\otimes\sigma_3),
\end{align}
and the expressions above can be rewritten as
\begin{align}
A_0^{(12)}&=-\frac{1}{2}\left(\frac{d^{2}}{d\rho_1^{2}}+\frac{d^{2}}{d\rho_2^{2}}\right)+\frac{1}{2}(\rho_1^{2}+\rho_2^{2})+\frac{k_1(k_1+\Sigma_{12})}{2\rho_1^{2}}+\frac{k_2(k_2+\Sigma_{34})}{2\rho_2^{2}},\\
A_\pm^{(12)}&=\ed\frac{1}{\sqrt{2}}\left[\gamma_1\left(\rho_1\mp\frac{d}{d\rho_1}\right)\mp i\gamma_2\frac{k_1}{\rho_1}\right]
             +\frac{1}{\sqrt{2}}\left[\gamma_3\left(\rho_2\mp\frac{d}{d\rho_2}\right)\mp i\gamma_4\frac{k_2}{\rho_2}\right],\label{eq:Apm12gamma}\\
P^{(12)}&=\eud\Sigma_{12}\Sigma_{34}.
\end{align}
We identify $A_0^{(12)}$ as the Hamiltonian of our two-dimensional spinorial model. Its energies are provided by the sum of two linear spectra:
\begin{align}
 n_1+n_2+\mu_1+\mu_2+1.
\end{align}
This spectrum is degenerate. We will now explain the degeneracies from the symmetries of the Hamiltonian $H_{12}=A_0^{(12)}$ and the algebra they form.

\subsection{Symmetries of $H_{12}=A_0^{(12)}$ and superintegrability}
The total Hamiltonian of the two-dimensional system is
\begin{align}
 H_{12}=\Delta(A_0)=A_0^{(1)}+A_0^{(2)}
\end{align}
where
\begin{align}
 A_0^{(1)}&=-\frac{1}{2}\left(\frac{d^{2}}{d\rho_1^{2}}\right)+\frac{1}{2}(\rho_1^{2})+\frac{k_1(k_1+\Sigma_{12})}{2\rho_1^{2}},\\
 A_0^{(2)}&=-\frac{1}{2}\left(\frac{d^{2}}{d\rho_2^{2}}\right)+\frac{1}{2}(\rho_2^{2})+\frac{k_2(k_2+\Sigma_{34})}{2\rho_2^{2}}.
\end{align}
Given the ladder operators \eqref{eq:osp12ap}--\eqref{eq:osp12am}, combinations of the form $A_{\pm}^{(i)}A_{\mp}^{(j)}$ will be symmetries of the total Hamiltonian $H_{12}$ (they are constants of motion that preserve the total energy of the system). We can form three independent such combinations (in addition to the Hamiltonian) that commute with $H_{12}$, thus showing that the model is superintegrable.\\[1em]
To identify the nature of the symmetry algebra, let
\begin{align}
 K_1=\frac{1}{4}\left(\{A_+^{(1)},A_-^{(1)}\}-\{A_+^{(1)},A_-^{(1)}\}\right)=\frac{1}{2}\left(A_0^{(1)}-A_0^{(2)}\right).
\end{align}
It is immediate to see that $K_1$ is a symmetry of $H_{12}$.

Another symmetry is $(A_-^{(1)}A_+^{(2)}-A_+^{(1)}A_-^{(2)})$; in essence, this is the total Casimir $Q^{(12)}=\Delta(Q)$ presented in \eqref{eq:Q12algebra}. This element obviously commutes with $H_{12}=\Delta(A_0)$ because the coproduct is an algebra homomorphism.

Note that both $\Sigma_{12}$ and $\Sigma_{34}$ are additional symmetries of $H_{12}$. A natural symmetry algebra generator is hence the total sCasimir $\Delta(S)$ which we shall denote by $K_2$:
\begin{align}
 K_2=\eud Q^{(12)}\Sigma_{12}\Sigma_{34}.
\end{align}
The commutation relations obeyed by the generators of the symmetry algebra are then seen to be the defining relations of the dual $-1$ Hahn algebra \cite{Genest2013}
\begin{align}\label{eq:dual-1Hahnalgebra}
\begin{aligned}{}
[K_1,K_2]&=K_3,\qquad\qquad [K_1,K_3]=K_2-(\eu Q^{(1)}\Sigma_{12}+\ed Q^{(2)}\Sigma_{34}-\tfrac{1}{2}),\\[0.5em]
[K_2,K_3]&=2K_3(\eu Q^{(1)}\Sigma_{12}+\ed Q^{(2)}\Sigma_{34})-4K_1(1-\eu Q^{(1)}\Sigma_{12}-\ed Q^{(2)}\Sigma_{34})\\
 &\hspace{14em}-2H_{12}(\eu Q^{(1)}\Sigma_{12}-\ed Q^{(2)}\Sigma_{34}),\\[0.5em]
 \{K_2,\Sigma_{12}\}&=2(\eu Q^{(1)}\Sigma_{12}+\ed Q^{(2)}\Sigma_{34}+\tfrac{1}{2})\Sigma_{12},\\
 \{K_2,\Sigma_{34}\}&=2(\eu Q^{(1)}\Sigma_{12}+\ed Q^{(2)}\Sigma_{34}+\tfrac{1}{2})\Sigma_{34},\\[.5em]
 0&=[K_1,\Sigma_{12}]=[K_1,\Sigma_{34}]=\{K_3,\Sigma_{12}\}=\{K_3,\Sigma_{34}\},
\end{aligned}
\end{align}
where $Q^{(1)}$, $Q^{(2)}$ and $H_{12}$ are central elements.
The algebra can also be recast in a more symmetric presentation by reabsorbing the $\epsilon_i$'s:
\begin{align}\label{eq:dual-1Hahnalgebrav2}
\begin{aligned}{}
[K_1,K_2]&=K_3,\qquad\qquad [K_1,K_3]=K_2-(S^{(1)}+S^{(2)}-\tfrac{1}{2}),\\[0.5em]
[K_2,K_3]&=2K_3(S^{(1)}+S^{(2)})+4K_1(S^{(1)}+S^{(2)}-1)-2H_{12}(S^{(1)}-S^{(2)}),\\[0.5em]
 \{K_2,P^{(1)}\}&=2(S^{(1)}+S^{(2)}+\tfrac{1}{2})P^{(1)},\\
 \{K_2,P^{(2)}\}&=2(S^{(1)}+S^{(2)}+\tfrac{1}{2})P^{(2)},\\[.5em]
 0&=[K_1,P^{(1)}]=[K_1,P^{(2)}]=\{K_3,P^{(1)}\}=\{K_3,P^{(2)}\},
\end{aligned}
\end{align}
This presentation emphasizes the role of the sCasimirs $S^{(i)}=Q^{(i)}P^{(i)}$ and makes the correspondence with the presentation in \cite{Genest2013} more explicit.

\subsection{An embedding of the dual Hahn algebra}
The dual Hahn algebra is connected with the Lie algebra $\su(1,1)$, which is the even subalgebra of $\osp(1|2)$. A natural question then arises: can the dual Hahn algebra be embedded in the $-1$ Hahn algebra? As will now be shown, the answer is affirmative.
\begin{align}\label{eq:osp12tosu11}
 J_{\pm}=\tfrac{1}{2}(A_{\pm})^{2},\qquad J_0=\tfrac{1}{2}A_0,
\end{align}
obey the defining relations of $\su(1,1)$:
\begin{align}
 [J_0,J_\pm]=\pm J_\pm,\qquad [J_+,J_-]=-2J_0.
\end{align}
The Casimir element of $\su(1,1)$ commutes with all $J_0$, $J_\pm$ generators and has the following expression in terms of the sCasimir $S=QP$ of $\osp(1|2)$:
\begin{align}
 C={J_0}^{2}-J_+J_--J_0=\tfrac{1}{4}\left(S^{2}+S-\tfrac{3}{4}\right).
\end{align}
The dual Hahn algebra appears when looking at the Clebsch-Gordan problem of $\su(1,1)$. Here is how this is realized in terms of a two-dimensional model.

Consider the addition of the two realizations \eqref{eq:ApmA0} of $\osp(1|2)$ and the corresponding two-dimensional $\su(1,1)$ model:
\begin{align}
 J_{\pm}^{(12)}&=\frac{1}{2}(A_{\pm}^{(12)})^{2}=\frac{1}{2}\left((A_{\pm}^{(1)})^{2}+(A_{\pm}^{(2)})^{2}\right),\\
 J_{0}^{(12)}&=\frac{1}{2}(A_{0}^{(12)})^{\phantom{2}}=\frac{1}{2}(A_{0}^{(1)}+A_{0}^{(2)}),\\
 C^{(12)}&=(J_0^{(12)})^{2}-J_+^{(12)}J_-^{(12)}-J_0^{(12)}.
\end{align}
Now, form the following two quantities that commute with the total Hamiltonian $J_0^{(12)}$:
\begin{align}
\begin{aligned}
 \cK_1&=\frac{1}{2}\left(J_0^{(1)}-J_0^{(2)}\right),\\
 \cK_2&=C^{(12)}=(J_0^{(12)})^{2}-J_+^{(12)}J_-^{(12)}-J_0^{(12)},
\end{aligned}
\end{align}
The relations obeyed by these elements are those of the dual Hahn algebra \cite{Granovskii1991,Granovskii1988,Frappat2019a}
\begin{align}
\begin{aligned}{}
 [\cK_1 ,\cK_2]&=\cK_3,\\
 [\cK_2,\cK_3]&=-2\{\cK_1,\cK_2\}+4J_0^{(12)}(C^{(1)}-C^{(2)}),\\
 [\cK_2,\cK_3]&=-2{\cK_1}^{2}-4\cK_2+2(J_0^{(12)})^{2}+4(C^{(1)}+C^{(2)}),
\end{aligned}
\end{align}
with central elements $\delta_1=4J_0^{(12)}(C^{(1)}-C^{(2)})$ and $\delta_2=2(J_0^{(12)})^{2}+4(C^{(1)}+C^{(2)})$. This explicitly shows the embedding of the dual Hahn algebra in the dual $-1$ Hahn algebra.

\section{Separated solutions}\label{sec:solutions}
In this section we shall study the wavefunctions of the two-dimensional system described by the Hamiltonian
\begin{align}
 H_{12}=-\frac{1}{2}\left(\frac{d^{2}}{d\rho_1^{2}}+\frac{d^{2}}{d\rho_2^{2}}\right)+\frac{1}{2}(\rho_1^{2}+\rho_2^{2})+\frac{k_1(k_1+\Sigma_{12})}{2\rho_1^{2}}+\frac{k_2(k_2+\Sigma_{34})}{2\rho_2^{2}}
\end{align}
in two different coordinates system. In the next section, the knowledge of the symmetry algebra will allow to obtain the overlaps between the wavefunctions in these two coordinate systems.

\subsection{Solutions in Cartesian coordinates}
The one-dimensional system obeys the following Schr\"odinger equation
\begin{align}\label{eq:1dschrod}
 H\psi=\frac{1}{2}\left(-\frac{d^{2}}{d\rho^{2}}+\rho^{2}+\frac{k(k-\sigma_3)}{\rho^{2}}\right)\psi=E\psi.
\end{align}
We give details in the Appendix \ref{sec:differential} on how this equation is solved.

The solutions $\psi_{m,k}(\rho)$ are more conveniently expressed in terms of the generalized Hermite polynomials $H_{m}^{k}(x)$ \cite{Chihara1978,Rosenblum1994,Rosler1998}. This family of polynomials is composed of two alternating sequences of generalized Laguerre polynomials:
\begin{align}\label{eq:genHermite}
 H_{2n+p}^{k}(x)=(-1)^{n}\sqrt{\frac{n!}{\Gamma(n+p+k+\frac{1}{2})}}~x^{p}~L_{n}^{(k-\frac{1}{2}+p)}(x^{2}),\qquad p\in\{0,1\}.
\end{align}
Identifying $p$ with $\frac{1-s}{2}$, where $s$ is the eigenvalue of $\sigma_3$, the solutions are presented as
\begin{align}
\begin{aligned}
 \psi_{m,k}(\rho)&=(-1)^{\lfloor\frac{m}{2}\rfloor}e^{-\rho^{2}/2}\rho^{k}\,H_{m}^{k}(\rho),\qquad\text{with}\quad m=2n+p,\\[0.5em]
 E_m&=m+k+\tfrac{1}{2},
\end{aligned}
\end{align}
and $\lfloor x\rfloor$ is the floor function. In braket notation, the Schr\"odinger equation reads
\begin{align}
 H\ket{m,k}=(m+k+\tfrac{1}{2})\ket{m,k},
\end{align}
with
\begin{align}\label{eq:spinorparity}
 \ket{m,k}=\begin{pmatrix} \psi_{m,k}\\0 \end{pmatrix} \qquad \text{if $m$ is even},\qquad\text{and}\qquad \ket{m,k}=\begin{pmatrix} 0\\\psi_{m,k} \end{pmatrix} \qquad \text{if $m$ is odd}.
\end{align}
One should note that we must have $k>\nicefrac{-1}{2}$ in order for the solutions to be normalizable and the energies to be non-negative. The action of the parity involution $P:\rho\mapsto-\rho$ on these wavefunctions is
\begin{align}
 P\,\psi_{m,k}(\rho)=\psi_{m,k}(-\rho)=(-1)^{k+m}\psi_{m,k}(\rho).
\end{align}
Since $P$ acts as $\epsilon(-1)^{m}$ on a given $\osp(1|2)$ eigenvector $\ket{m,k}$, this means that we should choose
\begin{align}
 \epsilon=(-1)^{k}.
\end{align}
Since the positive-discrete series of $\osp(1|2)$ in \eqref{eq:osp12a0}--\eqref{eq:actionP} is defined for $\epsilon=\pm1$, we impose $k\in\mathbb{N}$.

It can be checked (using the Laguerre polynomials contiguity and recurrence relations \cite{Rainville1960}) that $A_{\pm}$ realized as \eqref{eq:ApmA0} acts on the eigenstates $\ket{m,k}$ according to \eqref{eq:osp12ap}--\eqref{eq:osp12am}.
\begin{align}
 A_+\ket{m,k}=\sqrt{[m+1]_k}\ket{m+1,k},\qquad A_-\ket{m,k}=\sqrt{[m]_k}\ket{m-1,k}.
\end{align}
We now look at the solutions of the two-dimensional system. The Cartesian solutions are obtained by combining two one-dimensional problems. Let us denote the coupled eigenstates
\begin{align}
 \ket{m_1,k_1}\otimes\ket{m_2,k_2}=\ket{m_1,m_2,k_1,k_2},\qquad\quad m_i=2n_i+p_i.
\end{align}
These $\ket{m_1,m_2,k_1,k_2}$ are $4$-component spinors, whose entries depend on the parity of both $m_1$ and $m_2$ following \eqref{eq:spinorparity},
\begin{align}\hspace{-2em}
\begin{aligned}
 \ket{2n_1,2n_2,k_1,k_2}&=\begin{pmatrix} \psi_{2n_1,2n_2,k_1,k_2}\\0\\0\\0 \end{pmatrix},\\
 \ket{2n_1+1,2n_2,k_1,k_2}&=\begin{pmatrix} 0\\0\\\psi_{2n_1+1,2n_2,k_1,k_2}\\0 \end{pmatrix},
\end{aligned}\qquad
\begin{aligned}
 \ket{2n_1,2n_2+1,k_1,k_2}&=\begin{pmatrix} 0\\\psi_{2n_1,2n_2+1,k_1,k_2}\\0\\0 \end{pmatrix},\\
  \ket{2n_1+1,2n_2+1,k_1,k_2}&=\begin{pmatrix} 0\\0\\0\\\psi_{2n_1+1,2n_2+1,k_1,k_2} \end{pmatrix},
\end{aligned}
\end{align}
and the entries of these spinors are the product of two one-dimensional wavefunctions
\begin{align}
 \psi_{m_1,m_2,k_1,k_2}(\rho_1,\rho_2)=\psi_{m_1,k_1}(\rho_1)\psi_{m_2,k_2}(\rho_2).
\end{align}
The actions \eqref{eq:osp12a0}--\eqref{eq:actionP} extend naturally to the two-dimensional case. Moreover, from the action of $\Delta(P)=P_1P_2:(\rho_1,\rho_2)\mapsto(-\rho_1,-\rho_2)$ on $\psi_{m_1,m_2,k_1,k_2}(\rho_1,\rho_2)$, it is checked that $\eud=\eu\ed$ as expected. The energies can be presented as
\begin{align}\label{eq:cartesianenergiesm}
 E_{m_1,m_2}=(m_1+m_2)+(k_1+k_2)+1
\end{align}
and have additional degeneracies, as is seen from
\begin{align}
 m_1+m_2=\begin{cases}
          2(n_1+n_2)\phantom{+2}\qquad &\text{if}~(p_1,p_2)=(0,0),\\
          2(n_1+n_2)+1\qquad &\text{if}~(p_1,p_2)=(0,1),\\
          2(n_1+n_2)+1\qquad &\text{if}~(p_1,p_2)=(1,0),\\
          2(n_1+n_2)+2\quad &\text{if}~(p_1,p_2)=(1,1).
         \end{cases}
\end{align}
This two-fold degeneracy comes from the fact that the Hamiltonian \eqref{eq:H} is invariant under the exchange of $\rho_1\leftrightarrow\rho_2$, $k_1\leftrightarrow k_2$ and the internal spaces $2\leftrightarrow3$.

\subsection{Solutions in polar coordinates}
Superintegrable systems typically admit separation of variable in more than one coordinate system \cite{Miller2013}. This is the case here: one can also separate the solutions in polar coordinates. Write
\begin{align}
 \rho_1=r\cos\phi,\qquad \rho_2=r\sin\phi,
\end{align}
the Schr\"odinger equation takes the form
\begin{align}\label{eq:schrodpolar}
 H\Psi=\frac{1}{2}\left[-\frac{\partial^{2}}{\partial r^{2}}-\frac{1}{r}\frac{\partial}{\partial r}+r^{2}-\frac{1}{r^{2}}\left(\frac{\partial^{2}}{\partial\phi^{2}}-\frac{k_1(k_1-\sigma_3\otimes1)}{\cos^{2}\phi}-\frac{k_2(k_2-1\otimes\sigma_3)}{\sin^{2}\phi}\right)\right]\Psi=E\Psi.
\end{align}
To make the separation of variable in polar coordinates, write $\Psi(r,\phi)=R(r)\Phi(\phi)$. This yields the angular equation
\begin{align}\label{eq:polarangulari}
 \left[\frac{d^{2}}{d\phi^{2}}-\frac{k_1(k_1-\sigma_3\otimes1)}{\cos^{2}\phi}-\frac{k_2(k_2-1\otimes\sigma_3)}{\sin^{2}\phi}+m^{2}\right]\Phi=0
\end{align}
and the radial equation
\begin{align}
 \frac{d^{2}R}{dr^{2}}+\frac{1}{r}\frac{dR}{dr}+\left(-r^{2}-\frac{m^{2}}{r^{2}}+2E\right)R=0.
\end{align}
These equations are solved in Appendix \ref{sec:differential}.\\[1em]
The orthonormalized angular solution to \eqref{eq:polarangulari} is given in terms of Jacobi polynomials $J_{d}^{(\alpha,\beta)}(\phi)$:
\begin{align}
\begin{aligned}{}
 \Phi_{\ell,s_1,s_2}^{k_1,k_2}(\phi)&=\sqrt{\frac{2(2\ell+k_1+k_2)~(\ell-\frac{2-s_1-s_2}{4})!~\Gamma(\ell+k_1+k_2+\frac{2-s_1-s_2}{4})}{\Gamma(\ell+k_1+\frac{2-s_1+s_2}{4})~\Gamma(\ell+k_2+\frac{2+s_1-s_2}{4})}}\\
 &\qquad\times[\cos\phi]^{k_1+\frac{1-s_1}{2}}[\sin\phi]^{k_2+\frac{1-s_2}{2}}\,P_{\ell-\frac{2-s_1-s_2}{4}}^{(k_1-\frac{s_1}{2},k_2-\frac{s_2}{2})}(-\cos2\phi),\\[.5em]
 |m|&=2\ell+(k_1+k_2),\qquad s_1,s_2\in\{\pm1\}.
\end{aligned}
\end{align}
It is important to recall that in the above, $\ell\in\{0,1,2,\dots\}$ is a non-negative integer if $s_1s_2=1$ and $\ell\in\{\frac{1}{2},\frac{3}{2},\dots\}$ is a half-integer if $s_1s_2=-1$.
Also, the normalization condition is
\begin{align}
 \int_{0}^{\pi/2}d\phi\left[\Phi_{\ell,s_1,s_2}^{k_1,k_2}(\phi)\right]^{2}=1.
\end{align}
The solutions are defined in the first quadrant, which corresponds to the domain of definition of the Jacobi polynomials. In Section \ref{sec:reduction}, it will be seen that it is natural for the solutions to be given only in the first quadrant. The wavefunctions can be extended straightforwardly to the four quadrants.\\[1em]
Next, the solutions of the radial equation are expressed in terms of Laguerre polynomials like in the one-dimensional case:
\begin{align}
 R_{N'}(r)=\sqrt{\frac{2(N!)}{\Gamma(N'+|m|+1)}}r^{|m|}e^{-r^{2}/2}L_{N'}^{(|m|)}(r^{2}).
\end{align}
The orthonormalized solutions of the two-dimensional system in polar coordinates are then given by
\begin{align}
 \psi_{\ell,N',s_1,s_2}^{k_1,k_2}(r,\phi)=\Phi_{\ell,s_1,s_2}^{k_1,k_2}(\phi)R_{N'}(r)
\end{align}
The total energy of the two-dimensional system is
\begin{align}
\begin{aligned}
 E&=2N'+|m|+1=2(N'+\ell)+(k_1+k_2)+1,
\end{aligned}
\end{align}
which matches what was obtained for the Cartesian solutions in \eqref{eq:cartesianenergiesm} upon letting
\begin{align}
 2(N'+\ell)=m_1+m_2.
\end{align}
Equation \eqref{eq:schrodpolar} can alternatively be presented in the form
\begin{align}
 H\ket{\ell,N',s_1,s_2,k_1,k_2}=E\ket{\ell,N',s_1,s_2,k_1,k_2}
\end{align}
where the $\ket{\ell,N',s_1,s_2,k_1,k_2}$ are $4$-component spinors, whose entries depends on the value of $s_1$, $s_2$:
\begin{align}\label{eq:phipolar}
\begin{aligned}
 \ket{\ell,N',+,+,k_1,k_2}&=\begin{pmatrix} \psi_{\ell,N',+,+}^{k_1,k_2}\\0\\0\\0 \end{pmatrix},\\
 \ket{\ell,N',-,+,k_1,k_2}&=\begin{pmatrix} 0\\0\\\psi_{\ell,N',-,+}^{k_1,k_2}\\0 \end{pmatrix},
\end{aligned}\qquad\quad
\begin{aligned}
 \ket{\ell,N',+,-,k_1,k_2}&=\begin{pmatrix} 0\\\psi_{\ell,N',+,-}^{k_1,k_2}\\0\\0 \end{pmatrix},\\
  \ket{\ell,N',-,-,k_1,k_2}&=\begin{pmatrix} 0\\0\\0\\\psi_{\ell,N',-,-}^{k_1,k_2} \end{pmatrix}.
\end{aligned}
\end{align}

\section{Overlaps: the dual $-1$ Hahn polynomials}\label{sec:overlaps}
The goal is now to compute the overlaps between the wavefunctions arising from the separation of variables in Cartesian and polar coordinates and to relate these connection coefficients to the symmetry algebra exhibited previously in Section \ref{sec:symalgebra}.

\subsection{A basis diagonalizing $Q^{(12)}$}
The separation of variables in polar coordinates amounted to the diagonalization of the operator
\begin{align}
 \textbf{B}_\phi=-\frac{d^{2}}{d\phi^{2}}+\frac{k_1(k_1-\sigma_3\otimes1)}{\cos^{2}\phi}+\frac{k_2(k_2-1\otimes\sigma_3)}{\sin^{2}\phi}
\end{align}
We will now diagonalize $Q^{(12)}$. The reason for this choice is that the overlaps between the Cartesian eigenbasis and the eigenbasis of $Q^{(12)}$ are straightforward to obtain in light of our knowledge of the action of the ladder operators on an $\osp(1|2)$ irrep \eqref{eq:osp12a0}--\eqref{eq:actionP} and of that fact that the eigenstates of $Q^{(12)}$ can be obtained as a linear combination of the polar eigenstates.

In our realization
\begin{align}\label{eq:Q12inrealiz}
 Q^{(12)}=\left[-i\ed\frac{d}{d\phi}\Sigma_{13}-\ed k_1\frac{\sin\phi}{\cos\phi}\Sigma_{23}+\ed k_2\frac{\cos\phi}{\sin\phi}\Sigma_{14}+k_1\Sigma_{12}+k_2\Sigma_{34}-\frac{1}{2}\right]\eud\Sigma_{12}\Sigma_{34}.
\end{align}
Let us now fix the eigenvalue of $\Sigma_{12}\Sigma_{34}$ to be $\delta=\pm1$. We shall look for the eigenvectors of $Q^{(12)}$ such that
\begin{align}\label{eq:Q12diagbasis}
 Q^{(12)}\ket{F_{\delta}}=q_{\delta}\ket{F_{\delta}}.
\end{align}
Looking at the eigenvalue of $\Sigma_{34}$, which is either $\pm1$, we write
\begin{align}
 \ket{F_{\delta}}=\ket{f_{\delta}^{+}}+\ket{f_{\delta}^{-}},
\end{align}
with the labels $\pm$ chosen so that
\begin{align}
 \Sigma_{34}\ket{f_{\delta}^{\pm}}=\pm\ket{f_{\delta}^{\pm}}.
\end{align}

\subsubsection{The case $\delta=+1$}
In the case that $\delta=+1$ one has
\begin{align}
 \ket{F_{+}}=\begin{pmatrix} f_{+}^{-}\\0\\0\\f_{+}^{+} \end{pmatrix} = \ket{f_{+}^{-}}+\ket{f_{+}^{+}}
            = f_{+}^{-} \begin{pmatrix} 1\\0\\0\\0 \end{pmatrix}
             +f_{+}^{+} \begin{pmatrix} 0\\0\\0\\1 \end{pmatrix} = f_{+}^{-}\ket{+,-}_\Sigma+f_{+}^{+}\ket{+,+}_\Sigma.
\end{align}
One sees that $f_{+}^{+}$ is related to $f_{+}^{-}$ by virtue of \eqref{eq:Q12diagbasis}. Now, explicitly writing down \eqref{eq:Q12inrealiz} gives rise to two equations that must be simultaneously satisfied
\begin{align}
 \eu\left(\phantom{-}\frac{d}{d\phi}-k_1\frac{\sin\phi}{\cos\phi}+k_2\frac{\cos\phi}{\sin\phi}\right)f_{+}^{+}-\eud\left(k_1+k_2+\tfrac{1}{2}\right)f_{+}^{-}&=q_+f_{+}^{-},\label{eq:fpp}\\
 \eu\left(-\frac{d}{d\phi}-k_1\frac{\sin\phi}{\cos\phi}+k_2\frac{\cos\phi}{\sin\phi}\right)f_{+}^{-}+\eud\left(k_1+k_2-\tfrac{1}{2}\right)f_{+}^{+}&=q_+f_{+}^{+}.\label{eq:fpm}
\end{align}
Inserting \eqref{eq:fpp} into \eqref{eq:fpm}, denoting $\tilde{q}_{+}=q_++\tfrac{1}{2}\eud$ and recalling that $\eu^{2}=\ed^{2}=1$, we obtain the following second-order differential equation:
\begin{align}
 \frac{d^{2}f_{+}^{-}}{d\phi^{2}}-\left(\frac{k_1(k_1-1)}{\cos^{2}\phi}+\frac{k_2(k_2-1)}{\sin^{2}\phi}-\tilde{q}_+^{2}\right)f_{+}^{-}=0.
\end{align}
Comparing with \eqref{eq:polarangular}, it is easily seen that the solutions are given in terms of Jacobi polynomials (the exact normalization will be given below). We have
\begin{align}
\begin{aligned}
 f_{+}^{-}&=\sqrt{\frac{2~(\ell!)~(2\ell+k_1+k_2)~\Gamma(\ell+k_1+k_2)}{
\Gamma(\ell+k_1+\frac{1}{2})~\Gamma(\ell+k_2+\frac{1}{2})}}
(\cos\phi)^{k_1}(\sin\phi)^{k_2}P_{\ell}^{(k_1-\frac{1}{2},k_2-\frac{1}{2})}(-\cos2\phi)\\
 &=\Phi_{\ell,+,+}^{k_1,k_2}(\phi),
\end{aligned}
\end{align}
and the eigenvalues $\tilde{q}_{+}$ are
\begin{align}\label{eq:qplusvalue}
 \tilde{q}_{+}^{2}=(2\ell+k_1+k_2)^{2},\qquad\Longrightarrow\quad \tilde{q}_{+}=\pm(2\ell+k_1+k_2),\qquad \ell\in\mathbb{N}.
\end{align}
To obtain $f_{+}^{+}$, one could repeat what was done for $f_{+}^{-}$ and solve the resulting second-order differential equation. This would yield, up to some normalization, $f_{+}^{+}=\Phi_{\ell,-,-}^{k_1,k_2}(\phi)$. Since the relative normalization between $f_{+}^{-}$ and $f_{+}^{+}$ is crucial, we will use \eqref{eq:fpm} instead. First suppose $\ell>0$. We note that
\begin{align}
 (\cos\phi)^{k_1}(\sin\phi)^{k_2}\frac{d}{d\phi}(\cos\phi)^{-k_1}(\sin\phi)^{-k_2}=\frac{d}{d\phi}+k_1\frac{\sin\phi}{\cos\phi}-k_2\frac{\cos\phi}{\sin\phi}.
\end{align}
Then, it is straightforward to obtain
\begin{align}
\begin{aligned}
 f_{+}^{+}&=\frac{-\eu}{\tilde{q}_+-\eud(k_1+k_2)}\sqrt{\frac{2~\ell!~\Gamma(\ell+k_1+k_2)~(2\ell+k_1+k_2)}{\Gamma(\ell+k_1+\frac{1}{2})~\Gamma(\ell+k_2+\frac{1}{2})}}\\
 &\qquad\times(\cos\phi)^{k_1}(\sin\phi)^{k_2}~\frac{d}{d\phi}P_{\ell}^{(k_1-\frac{1}{2},k_2-\frac{1}{2})}(-\cos2\phi),
\end{aligned}
\end{align}
and using \cite{Rainville1960}
\begin{align}\label{eq:derivativePn}
 \frac{d}{dx}P_{n}^{(\alpha,\beta)}(x)=\frac{n+\alpha+\beta+1}{2}P_{n-1}^{(\alpha+1,\beta+1)}(x),
\end{align}
one is led to
\begin{align}
\begin{aligned}
 f_{+}^{+}&=\frac{-2\eu\sqrt{\ell(\ell+k_1+k_2)}}{\tilde{q}_+-\eud(k_1+k_2)}\sqrt{\frac{2~(\ell-1)!~\Gamma(\ell+k_1+k_2+1)~(2\ell+k_1+k_2)}{\Gamma(\ell+k_1+\frac{1}{2})~\Gamma(\ell+k_2+\frac{1}{2})}}\\
 &\qquad\times(\cos\phi)^{k_1+1}(\sin\phi)^{k_2+1}~P_{\ell-1}^{(k_1+\frac{1}{2},k_2+\frac{1}{2})}(-\cos2\phi)\\
 &=\frac{-2\eu\sqrt{\ell(\ell+k_1+k_2)}}{\tilde{q}_+-\eud(k_1+k_2)}\Phi_{\ell,-,-}^{k_1,k_2}(\phi).
\end{aligned}
\end{align}
Recall that $\tilde{q}_+$ has two possible expressions $\pm(2\ell+k_1+k_2)$. Consider each case in turn. This gives two solutions, corresponding to each sign. The full orthonormalized solutions to \eqref{eq:fpp}--\eqref{eq:fpm} are then found to be:
\begin{align}
 \hspace{-2em}\ket{F_{\ell,+,+}}&=\sqrt{\frac{\ell+\frac{1-\eud}{2}(k_1+k_2)}{2\ell+k_1+k_2}}\ket{\Phi_\ell;+,-}-\eu\sqrt{\frac{\ell+\frac{1+\eud}{2}(k_1+k_2)}{2\ell+k_1+k_2}}\ket{\Phi_\ell;+,+},\\
 \hspace{-2em}\ket{F_{\ell,+,-}}&=\sqrt{\frac{\ell+\frac{1+\eud}{2}(k_1+k_2)}{2\ell+k_1+k_2}}\ket{\Phi_\ell;+,-}+\eu\sqrt{\frac{\ell+\frac{1-\eud}{2}(k_1+k_2)}{2\ell+k_1+k_2}}\ket{\Phi_\ell;+,+},
\end{align}
with
\begin{align}
 \ket{\Phi_\ell;+,-}=\Phi_{\ell,+,+}^{k_1,k_2}(\phi)\ket{+,-}_\Sigma,\qquad \ket{\Phi_\ell;+,+}=\Phi_{\ell,-,-}^{k_1,k_2}(\phi)\ket{+,+}_\Sigma.
\end{align}
These solutions indeed diagonalize $Q^{(12)}$
\begin{align}
 Q^{(12)}\ket{F_{\ell,+,\pm}}=(\pm|\tilde{q}_+|-\tfrac{1}{2}\eud)~\ket{F_{\ell,+,\pm}}=\pm(2\ell+k_1+k_2\mp\tfrac{1}{2}\eud)\ket{F_{\ell,+,\pm}}.
\end{align}
In the case where $\ell=0$, $\Phi_{0,-,-}^{k_1,k_2}(\phi)$ vanishes and there exists only a single eigenstate of $Q^{(12)}$, whose eigenvalue equation is
\begin{align}
 Q^{(12)}\ket{F_{0,+,-\eud}}=-\eud(k_1+k_2+\tfrac{1}{2}).
\end{align}
Let us now define the eigenvectors $\ket{q_z}$, with $z\in\mathbb{N}$:
\begin{align}\label{eq:qnpu}
\begin{aligned}
 \ket{q_{2\ell}}=\ket{F_{\ell,+,-\eud}},\qquad \ket{q_{2\ell-1}}=\ket{F_{\ell,+,\phantom{-}\eud}},\qquad\quad\ell\in\mathbb{N}.
\end{aligned}
\end{align}
The spectrum of $Q^{(12)}$ can then be repackaged in a single expression
\begin{align}
 Q^{(12)}\ket{q_z}=q_z\ket{q_z},\qquad\quad q_{z}=\eud(-1)^{z+1}(z+k_1+k_2+\tfrac{1}{2}),\qquad z\in\mathbb{N}.
\end{align}

\subsubsection{The case $\delta=-1$}
We repeat the analysis for the case $\delta=-1$. One looks for eigenvectors
\begin{align}
 \ket{F_{-}}=\begin{pmatrix} 0\\f_{-}^{+}\\f_{-}^{-}\\0 \end{pmatrix} = \ket{f_{-}^{-}}+\ket{f_{-}^{+}}
            = f_{-}^{-} \begin{pmatrix} 0\\0\\1\\0 \end{pmatrix}
             +f_{-}^{+} \begin{pmatrix} 0\\1\\0\\0 \end{pmatrix} = f_{-}^{-}\ket{-,-}_\Sigma+f_{-}^{+}\ket{-,+}_\Sigma.
\end{align}
Writing down \eqref{eq:Q12diagbasis} in the realization leads to the following equations:
\begin{align}
 \eu\left(\phantom{-}\frac{d}{d\phi}-k_1\frac{\sin\phi}{\cos\phi}-k_2\frac{\cos\phi}{\sin\phi}\right)f_{-}^{-}+\eud\left(k_1-k_2+\tfrac{1}{2}\right)f_{-}^{+}&=q_-f_{-}^{+},\label{eq:fmm}\\
 \eu\left(-\frac{d}{d\phi}-k_1\frac{\sin\phi}{\cos\phi}-k_2\frac{\cos\phi}{\sin\phi}\right)f_{-}^{+}-\eud\left(k_1-k_2-\tfrac{1}{2}\right)f_{-}^{-}&=q_-f_{-}^{-}.\label{eq:fmp}
\end{align}
Similarly, substituting \eqref{eq:fmm} into \eqref{eq:fmp}, one obtains
\begin{align}
 \frac{d^{2}f_{-}^{-}}{d\phi^{2}}-\left(\frac{k_1(k_1+1)}{\cos^{2}\phi}+\frac{k_2(k_2-1)}{\sin^{2}\phi}-\tilde{q}_-^{2}\right)f_{-}^{-}&=0,
\end{align}
with $\tilde{q}_-=q_--\frac{1}{2}\eud$.
Comparing with \eqref{eq:polarangular}, it is easily seen that the solutions are again given in terms of Jacobi polynomials:
\begin{align}
 f_{-}^{-}=\Phi_{\ell,-,+}^{k_1,k_2}(\phi),\qquad \ell\in\{\tfrac{1}{2},\tfrac{3}{2},\dots\}
\end{align}
and the eigenvalues are found to be
\begin{align}\label{eq:qminusvalue}
 \tilde{q}_{-}^{2}=(2\ell+k_1+k_2)^{2},\qquad\Longrightarrow\quad \tilde{q}_{-}=\pm(2\ell+k_1+k_2).
\end{align}
We are interested in the relative normalization between $f_{-}^{-}$ and $f_{-}^{+}$, so we use \eqref{eq:fmm}. It will be useful to call upon the identity
\begin{align}
 \frac{d}{dz}P_{n}^{(\alpha,\beta)}(z)=\frac{\alpha+n}{z-1}P_{n}^{(\alpha-1,\beta+1)}(z)-\frac{\alpha}{z-1}P_{n}^{(\alpha,\beta)}(z),
\end{align}
which is obtained by combining the Laguerre mixed relations \cite[p.~$264$, equations (9)$-$(17)]{Rainville1960} with \eqref{eq:derivativePn}.

Treating the two possible values of $\tilde{q}_{-}$, we obtain two solutions. Finally, the orthonormalized basis is obtained:
\begin{align}
 \hspace{-2em}\ket{F_{\ell,-,+}}&=\sqrt{\frac{\ell+\frac{1-\eud}{2}k_1+\frac{1+\eud}{2}k_2}{2\ell+k_1+k_2}}\ket{\Phi_\ell;-,-}-\eu\sqrt{\frac{\ell+\frac{1+\eud}{2}k_1+\frac{1-\eud}{2}k_2}{2\ell+k_1+k_2}}\ket{\Phi_\ell;-,+},\\
 \hspace{-2em}\ket{F_{\ell,-,-}}&=\sqrt{\frac{\ell+\frac{1+\eud}{2}k_1+\frac{1-\eud}{2}k_2}{2\ell+k_1+k_2}}\ket{\Phi_\ell;-,-}+\eu\sqrt{\frac{\ell+\frac{1-\eud}{2}k_1+\frac{1+\eud}{2}k_2}{2\ell+k_1+k_2}}\ket{\Phi_\ell;-,+},
\end{align}
with
\begin{align}
 \ket{\Phi_\ell;-,-}=\Phi_{\ell,-,+}^{k_1,k_2}(\phi)\ket{-,-}_\Sigma,\qquad \ket{\Phi_\ell;-,+}=\Phi_{\ell,+,-}^{k_1,k_2}(\phi)\ket{-,+}_\Sigma.
\end{align}
In this basis,
\begin{align}
 Q^{(12)}\ket{F_{-,\pm}}=(\pm|\tilde{q}_-|+\tfrac{1}{2}\eud)~\ket{F_{-,\pm}}=\pm(2\ell+k_1+k_2\pm\tfrac{1}{2}\eud)\ket{F_{-,\pm}}.
\end{align}
Defining the eigenvectors $\ket{q_z}$ for $z\in\mathbb{N}$ and $\ell\in\mathbb{N}$ as
\begin{align}\label{eq:qnmu}
 \ket{q_{2\ell}}=\ket{F_{\ell,+,-\eud}},\qquad \ket{q_{2\ell+1}}=\ket{F_{\ell,-,\eud}},\qquad\quad\ell\in\mathbb{N},
\end{align}
the spectrum of $Q^{(12)}$ can be presented in a single expression
\begin{align}
 Q^{(12)}\ket{q_z}=q_z\ket{q_z},\qquad\quad q_{z}=\eud(-1)^{z+1}(z+k_1+k_2+\tfrac{1}{2}),\qquad z\in\mathbb{N}.
\end{align}

\subsection{Overlaps with the Cartesian basis and the dual $-1$ Hahn polynomials}
It is clear that the overlaps between eigenstates with different energies will vanish. Let us then consider cases where $m_1+m_2=2(N'+\ell)$, i.e. cases where the energies of the eigenstates in Cartesian coordinates and polar coordinates are equal.

The overlaps between the eigenvectors $\ket{q_z}$ diagonalizing $Q^{(12)}$ and the Cartesian eigenvectors $\ket{m_1,m_2,k_1,k_2}\equiv\ket{m_1;m_2}$ are:
\begin{align}
 \Braket{q_{z}}{Q^{(12)}}{m_1;m_2}
 =\Braket{q_{z}}{(A_-^{(1)}A_+^{(2)}-A_+^{(1)}A_-^{(2)})P^{(1)}+Q^{(1)}P^{(2)}+Q^{(2)}P^{(1)}-\tfrac{1}{2}P^{(1)}P^{(2)}}{m_1;m_2}.
\end{align}
Using the actions given in \eqref{eq:osp12a0}--\eqref{eq:actionP} and defining $\braket{q_z}{m_1;m_2}=M_{m_1,m_2}$ yields
\begin{align}
\begin{aligned}
 &q_{z}~M_{m_1,m_2}=-\eud\left(k_1(-1)^{m_2}+k_2(-1)^{m_1}+\tfrac{1}{2}(-1)^{m_1+m_2}\right)M_{m_1,m_2}\\
 &+\eu(-1)^{m_1}\left(\sqrt{[m_1]_{k_1}[m_2+1]_{k_2}}M_{m_1-1,m_2+1}-\sqrt{[m_1+1]_{k_1}[m_2]_{k_2}}M_{m_1+1,m_2-1}\right).
\end{aligned}
\end{align}
Making the change of variables $N=m_1+m_2$,\quad $m=m_1$ and writing
\begin{align}
\begin{aligned}
 M_{m_1,m_2}&=\left(\frac{\ed}{2}\right)^{m}(-1)^{\frac{m(m+2N+1)}{2}}\left(\prod_{p_1=1}^{m}\prod_{p_2=1}^{N-m}\sqrt{\frac{[p_2]_{k_2}}{[p_1]_{k_1}}}~\right)\cN_{0,N}~\cN_{m,N},
\end{aligned}
\end{align}
we obtain a monic three term recurrence relation for the matrix elements:
\begin{align}\label{eq:TTRRgen}
\begin{aligned}
 (-1)^{N}[2\eud\,q_{z}]\cN_{m,N}=\cN_{m+1,N}&+[2(-1)^{m+1}(k_1+(-1)^{N}k_2)-1]\cN_{m,N}\\
 &+4[m]_{k_1}[N-m+1]_{k_2}~\cN_{m-1,N}.
\end{aligned}
\end{align}
A quick look at \eqref{eq:TTRRHahnmu} shows us that these matrix elements $\cN_{m,N}$ are dual $-1$ Hahn polynomials $P_{m}(x_{z};k_1,k_2,N)$ in the variable
\begin{align}
 x_{z}=(-1)^{N+z+1}(2z+2k_1+2k_2+1),\qquad \text{with\quad $z\in\mathbb{N}$}.
\end{align}
We then have the desired expression for $M_{m_1,m_2}=\braket{q_z}{m_1;m_2}$ up to a normalization, which can be determined using the orthonormality of the two bases:
\begin{align}\label{eq:orthooverlap}
 \delta_{m\bar m}=\sum_{q_z}\braket{\bar m,N-\bar m}{q_z}\braket{q_z}{m,N-m}.
\end{align}
If $N$ is odd, \eqref{eq:magicalgrid} tells us that we have $x_z=x_s$ by taking $z=s\in\{0,1,\dots,N\}$ and it follows that
\begin{align}\label{eq:overlapodd}
 \braket{q_z}{m,N-m}=P_{m}(x_{z};k_1,k_2,N)\sqrt{\frac{w_z(k_1,k_2,N)}{\nu_m(k_1,k_2,N)}}.
\end{align}
If $N$ is even, we recover $x_z=x_s$ by taking $z=N-s\in\{0,1,\dots,N\}$ and hence
\begin{align}\label{eq:overlapeven}
 \braket{q_z}{m,N-m}=P_{m}(x_{z};k_1,k_2,N)\sqrt{\frac{w_{N-z}(k_1,k_2,N)}{\nu_m(k_1,k_2,N)}}.
\end{align}
It is then simple to reexpress the eigenvectors $\ket{q_z}$ diagonalizing $Q^{(12)}$ in terms of the polar eigenvectors $\ket{\ell,N',s_1,s_2,k_1,k_2}$. Indeed, from the definitions of $\ket{q_z}$ in terms of $\ket{\Phi_\ell;s_1,s_2}$ in \eqref{eq:qnpu} and \eqref{eq:qnmu} as well as \eqref{eq:phipolar}, the relations are easily inversed.
From there, one can reexpress the overlaps between the polar and Cartesian eigenvectors as a linear combination of dual $-1$ Hahn polynomials following the results in \eqref{eq:overlapodd}--\eqref{eq:overlapeven}.

\section{Dimensional reduction}\label{sec:reduction}
The spinorial model obtained in Section \ref{sec:model} can furthermore be derived through a dimensional reduction procedure.

We start with a system of $4$ uncoupled standard harmonic oscillator acting on $4$ dimensional space, described by the Hamiltonian
\begin{align}\label{eq:Hinit}
 \tilde{H}=\sum_{i=1}^{4} a_i^{\dagger}a_i^{\phantom{\dagger}}+2\BLUE{\mathbb{I}_4},
\end{align}
where $a_i^{\phantom{\dagger}}$, $a_i^{\dagger}$ are the usual annihilation/creation operators
\begin{align}
 a_i^{\phantom{\dagger}}=\frac{1}{\sqrt{2}}\left(x_i+\frac{\partial}{\partial x_i}\right)\BLUE{\mathbb{I}_4},\qquad
 a_i^{\dagger}=\frac{1}{\sqrt{2}}\left(x_i-\frac{\partial}{\partial x_i}\right)\BLUE{\mathbb{I}_4},\qquad i=1,\dots,4,
\end{align}
satisfying $[a_i^{\dagger},a_j^{\phantom{\dagger}}]=\delta_{ij}\BLUE{\mathbb{I}_4}$ and $\BLUE{\mathbb{I}_4}$ is the $4$-dimensional identity matrix. One can now introduce the cylindrical coordinates
\begin{align}
\begin{aligned}
 x_1=\rho_1\cos\theta_1,\\
 x_2=\rho_1\sin\theta_1,
\end{aligned}\qquad
\begin{aligned}
 x_3=\rho_2\cos\theta_2,\\
 x_4=\rho_2\sin\theta_2.
\end{aligned}
\end{align}
The Hamiltonian \eqref{eq:Hinit} is rewritten as
\begin{align}
 \tilde{H}=\frac{1}{2}\sum_{i=1}^{2}\left(\rho_i^{2}-\frac{\partial^{2}}{\partial\rho_i^{2}}-\frac{1}{\rho_i}\frac{\partial}{\partial\rho_i}-\frac{1}{\rho_i^{2}}\frac{\partial^{2}}{\partial\theta_i^{2}}\right)\BLUE{\mathbb{I}_4}.
\end{align}
We can now effect the gauge transformation $\chi$ on the radii to get rid of the $\frac{1}{\rho_i}\frac{\partial}{\partial\rho_i}$ term:
\begin{align}\label{eq:gaugetrans}
 \chi(~\cdot~)=(\rho_1\rho_2)^{1/2}(~\cdot~)(\rho_1\rho_2)^{-1/2}.
\end{align}
This yields
\begin{align}\label{eq:radiigaugedHam}
 \chi(\tilde{H})=H=\frac{1}{2}\sum_{i=1}^{2}\left(\rho_i^{2}-\frac{\partial^{2}}{\partial\rho_i^{2}}-\frac{\nicefrac{1}{4}}{\rho_i^{2}}-\frac{1}{\rho_i^{2}}\frac{\partial^{2}}{\partial\theta_i^{2}}\right)\BLUE{\mathbb{I}_4}.
\end{align}
Owing to cylindrical symmetry, it is possible to set values for the spinorial angular momenta
\begin{align}\label{eq:J12J34fix}
\begin{aligned}
 J_{12}&=-i\frac{\partial}{\partial\theta_1}+\frac{1}{2}\Sigma_{12}=-k_1,\\
 J_{34}&=-i\frac{\partial}{\partial\theta_2}+\frac{1}{2}\Sigma_{34}=-k_2,
\end{aligned}
\end{align}
and the desired Hamiltonian $H_{12}$ is obtained:
\begin{align}\label{eq:reducedHam}
 H_{12}=\frac{1}{2}\left[-\left(\frac{d^{2}}{d\rho_1^{2}}+\frac{d^{2}}{d\rho_2^{2}}\right)+(\rho_1^{2}+\rho_2^{2})+\frac{k_1(k_1+\Sigma_{12})}{\rho_1^{2}}+\frac{k_2(k_2+\Sigma_{34})}{\rho_2^{2}}\right]\BLUE{\mathbb{I}_4}.
\end{align}
Recall that the solutions obtained in Section \ref{sec:solutions} were naturally defined for the first quadrant only, that is $\phi\in[0,\tfrac{\pi}{2}]$. It is now clear from the dimensional reduction procedure that this has to be the case. The reduced system coordinates $\rho_1$, $\rho_2$ are radial coordinates taking values in $\mathbb{R}^{+}$, and those positive coordinates are precisely those that define the first quadrant of the plane.

One may wonder if this dimensional reduction procedure can be carried out for all $\osp(1|2)$ generators (that is, for the $A_{\pm}^{(12)}$ as well). The answer is affirmative, but there is some refinement needed: one needs to perform an additional gauge transformation on the spinorial space. This was carried out in \cite{Schluter1983,Gaboriaud2018b}.

The reason for the need of an additional gauge transformation in spin space is that the gamma matrices appearing in the expressions \eqref{eq:Apm12gamma} of the $A_\pm^{(12)}$ acquire an angular dependency when one passes to cylindrical coordinates. The spinorial gauge transformation is meant to ``rotate out'' the angular dependency so that the procedure described in \eqref{eq:gaugetrans}--\eqref{eq:reducedHam} can be similarly applied.\\[.5em]

\noindent The novel aspects of the superconformal system presented here come from the presence of internal degrees of freedom whose non-trivial ties arise in dimensional reduction by fixing the spinorial angular momentum. Comparison with the situation where only the spatial angular momentum is fixed is instructive.  In case we set
\begin{align}
\begin{aligned}{}
 L_{12}&=-i\frac{\partial}{\partial\theta_1}=-\kappa_1,\\
 L_{34}&=-i\frac{\partial}{\partial\theta_2}=-\kappa_2,
\end{aligned}
\end{align}
the Hamiltonian \eqref{eq:radiigaugedHam} reduces to
\begin{align}
 H'_{12}=\frac{1}{2}\left[-\left(\frac{d^{2}}{d\rho_1^{2}}+\frac{d^{2}}{d\rho_2^{2}}\right)+(\rho_1^{2}+\rho_2^{2})+\frac{\kappa_1^{2}-\frac{1}{4}}{\rho_1^{2}}+\frac{\kappa_2^{2}-\frac{1}{4}}{\rho_2^{2}}\right]\BLUE{\mathbb{I}_4}.
\end{align}
This is in effect a system of two singular oscillators, and such oscillators are associated to the algebra $\mathfrak{su}(1,1)$ instead of $\osp(1|2)$. These have been examined in detail and shown to be superintegrable in \cite{Rodriguez2008}. The fact that many four-dimensional oscillators were a priori considered is immaterial under this reduction process.

However, if one fixes the spinorial angular momentum \eqref{eq:J12J34fix}, the reduction effectively couples the internal degrees of freedom. This is the origin of the relations \eqref{eq:conditionsk}, which connect the pure angular momentum (and thus the parity of the wavefunctions) with the spin (and thus the index of the components of the spinors).

\section{Conclusion}\label{sec:conclusion}
This paper has introduced a superconformal system with internal degrees of freedom in two dimensions that is superintegrable and that has the dual $-1$ Hahn algebra as its symmetry algebra. This model has been obtained by combining two spinorial realizations of the superalgebra $\osp(1|2)$ and identifying the Hamiltonian as the resulting Cartan generator.

What about combining more than two representations of $\osp(1|2)$?

It is known \cite{Genest2014a,Genest2014} that the generic superintegrable model on the two-sphere \cite{Kalnins1996} is obtained in a similar spirit by combining three realizations of $\su(1,1)$. In this case the Hamiltonian is taken to be the total Casimir element. A two-dimensional system is obtained because the norm of the radius vector is conserved. The constants of motion correspond to the intermediate Casimir operators which generate the symmetry algebra known under the name of Racah \cite{Kalnins2007,Genest2014a,Genest2014}. All other scalar second-order superintegrable models in two-dimensions can be obtained as special cases or contractions \cite{Kalnins2013} of this generic model.

Three parabosonic realizations of $\osp(1|2)$ have been similarly \cite{Genest2014f} combined to obtain a superintegrable model with reflections on the $2$-sphere that has the Bannai--Ito algebra as symmetry algebra. Here again the Hamiltonian is related (quadratically) to the Casimir operator of the underlying superalgebra. It has been shown \cite{Genest2014f} that the superintegrable Dunkl oscillator in two dimensions can in fact be obtained as a contraction of this Bannai--Ito invariant model on $S^{2}$.

These observations suggest that it would be relevant to combine three spinorial representations of $\osp(1|2)$ like the ones considered here to construct a model on $S^{2}$, without reflections, that should hence have by construction the Bannai--Ito algebra as its symmetry algebra. One would a priori expect the model presented here to be a contraction of the above. This raises interesting questions. One issue is that the number of degrees of freedom associated to combining two and three $\osp(1|2)$ representations differs from the start; another has to do with the fact that the dual $-1$ Hahn algebra is known to be a contraction of the algebra of the complementary Bannai--Ito polynomials \cite{Genest2013d} which is quite different from the Bannai--Ito one. Sorting this out should prove enlightening.

Now adding three spinorial realizations of $\osp(1|2)$ and taking as done here the Hamiltonian to be the total Cartan generator will yield a superintegrable singular oscillator with internal degrees of freedom in three dimensions. This has been performed with the parabosonic realizations to obtain the superintegrable Dunkl oscillator in three dimensions with an invariance algebra called the Schwinger-Dunkl algebra $sd(3)$ that extends $\su(3)$. We may thus expect a similar outcome in the case with internal degrees of freedom. While this has not been established, we could anticipate that the symmetry algebra, likely $sd(3)$, is isomorphic to the algebra of the rank $2$ dual $-1$ Hahn algebra. This would be the algebra associated to the bivariate or two-variable dual $-1$ Hahn polynomials that have not been characterized so far. While the bivariate Bannai--Ito polynomials have recently been introduced and studied \cite{Lemay2018}, this is not the case for the bivariate complementary Bannai--Ito polynomials from which the bivariate dual $-1$ Hahn polynomials should descend. With respect to contractions, these three-dimensional singular oscillators should relate to systems on the three sphere obtained by considering the addition of four realizations of $\osp(1|2)$ (see \cite{DeBie2016b} in this connection).

In another register, we wish to point out that one may use the $R$-matrix approach to arrive \cite{Harnad2004} at the generic superintegrable model on $S^{2}$ and construct its constants of motion. One proceeds via dimensional reduction with a Lax matrix that involves three $\su(1,1)$ elements. It has been shown recently \cite{Crampe2019a} that the universal $R$-matrix of $\osp(1|2)$ plays a central role in the description of the Bannai--Ito algebra. It should prove interesting to explore how this general formalism of integrable systems applies to the description of the superintegrable models with internal degrees of freedom that we have been discussing.

We thus observe that the superintegrable model introduced here presents itself as a nice basis to examine some of the various questions we have pointed out that pertains generally to the understanding of the algebras of Askey-Wilson type and their applications. We hope to follow up with these matters in the near future.

\subsection*{Acknowledgments}
While this research was conducted, PAB held an Undergraduate Student Research Award (USRA) from the Natural Sciences and Engineering Research Council of Canada (NSERC). JG holds an Alexander-Graham-Bell scholarship from the NSERC. The research of LV is supported in part by a Discovery Grant from NSERC.

\newpage

\appendix
\section{The dual $-1$ Hahn polynomials}\label{sec:dualHahn}
Here are a few useful definitions and properties of the dual $-1$ Hahn polynomials
\cite{Tsujimoto2013,Genest2013}, which have been introduced as a $q\to1$ limit of the dual
$q$-Hahn polynomials \cite{Koekoek2010}.

We denote the monic dual $-1$ Hahn polynomials $P_n(x;\xi,\zeta, N)$, where the parameters
$\xi,\zeta>-\tfrac{1}{2}$ and $N$ is an integer. These polynomials satisfy a three-term
recurrence relation
\begin{align}\label{eq:TTRRHahnmu}
 xP_n(x)=P_{n+1}(x) + [2(-1)^{n+1}(\xi+(-1)^N\zeta)-1]~P_n(x)
 + 4[n]_{\xi} [N-n+1]_{\zeta}~P_{n-1}(x)
\end{align}
Note that the factors are chosen for consistency with the definitions in references
\cite{Tsujimoto2013,Genest2013}.

Recall that the hypergeometric series $_rF_s$ is defined by
\begin{align}\label{eqn_hypergeometric}
 \pFq{r}{s}{a_1,\cdots,a_r}{b_1,\cdots,b_s}{z}
 = \sum_{k=0}^{\infty} \frac{(a_1)_k\cdots(a_r)_k}{(b_1)_k\cdots(b_s)_k} {\frac{z^k}{k!}}
\end{align}
with $(c)_k = c(c+1)\cdots(c+k-1)$ the Pochhammer symbol. The dual $-1$ Hahn polynomials
can be expressed as a generalized hypergeometric truncating series.\\[1em]
For $N$ even, denote $\delta=-\tfrac{1}{2}(\xi+\zeta+N)$; the expressions are
\begin{align}
 \hspace{-2em}
 P_{2n}(x)&=\phantom{(x+ 2\xi+2\zeta+1)}2^{4n}\left(\phantom{1}-\tfrac{N}{2}\right)_{n}
 \left(\tfrac{1}{2}-\tfrac{N}{2}-\zeta\right)_{n}~
 \pFq{3}{2}{-n,~\delta+\frac{1+x}{4},~\delta-\frac{1+x}{4}}
 {\phantom{1}-\frac{N}{2},~-\frac{N}{2}+\frac{1}{2}-\zeta}{1},\\
 \hspace{-2em}
 P_{2n+1}(x)&=(x+ 2\xi+2\zeta+1)2^{4n}\left(1-\tfrac{N}{2}\right)_{n}
 \left(\tfrac{1}{2}-\tfrac{N}{2}-\zeta\right)_{n}~
 \pFq{3}{2}{-n,~\delta+\frac{1+x}{4},~\delta-\frac{1+x}{4}}
 {1-\frac{N}{2},~-\frac{N}{2}+\frac{1}{2}-\zeta}{1}.
\end{align}
For $N$ odd, denote $\eta=\tfrac{1}{2}(\xi+\zeta+1)$; the expressions are
\begin{align}
 \hspace{-2em}
 P_{2n}(x)&=\phantom{\left(x+2\xi-2\zeta+1\right)}2^{4n}
 \left(\tfrac{1-N}{2}\right)_{n}\left(\xi+\tfrac{1}{2}\right)_{n}~
 \pFq{3}{2}{-n,~\eta+\frac{1+x}{4},~\eta-\frac{1+x}{4}}{\frac{1-N}{2},~\xi+\frac{1}{2}}{1},\\
 \hspace{-2em}
 P_{2n+1}(x)&=\left(x+2\xi-2\zeta+1\right)2^{4n}\left(\tfrac{1-N}{2}\right)_{n}
 \left(\xi+\tfrac{3}{2}\right)_{n}~
 \pFq{3}{2}{-n,~\eta+\frac{1+x}{4},~\eta-\frac{1+x}{4}}{\frac{1-N}{2},~\xi+\frac{3}{2}}{1}.
\end{align}
These polynomials obey an orthogonality relation of the form
\begin{align}
 \sum_{s=0}^N w_s(\xi,\zeta,N) P_n(x_s;\xi,\zeta,N) P_m(x_s;\xi,\zeta,N)
 = \nu_n (\xi,\zeta,N) \delta_{n,m}
\end{align}
on the grid points
\begin{align}\label{eq:magicalgrid}
 x_s=\quad
 \begin{cases}
  (-1)^{s}(2s-2\xi-2\zeta-2N-1)\qquad&\text{$N$ even},\\
  (-1)^{s}(2s+2\xi+2\zeta+1)&\text{$N$ odd}.\\
 \end{cases}
\end{align}
The weights are given by
\begin{align}
 \hspace{-2em}
 w_{2m+j}(\xi,\zeta,N)=\
 \begin{cases}
  \frac{(-1)^{m}\left(-\frac{N}{2}\right)_{m+j}}{m!}
\frac{\left(\frac{1-N}{2}-\zeta\right)_{m}}{\left(\frac{1-N}{2}-\xi\right)_{m}}
\frac{\left(-N-\xi-\zeta\right)_{m\phantom{+j}}}{\left(-\frac{N}{2}-\xi-\zeta
\right)_{m+j}}\qquad&\text{$N$ even},\\[1em]
  \frac{(-1)^{m}\left(\frac{1-N}{2}\right)_{m}}{m!}
\frac{\left(\xi+\frac{1}{2}\right)_{m+j}}{\left(\zeta+\frac{1}{2}\right)_{m+j}}
\frac{\left(1+\xi+\zeta\right)_{m}}{\left(\frac{1}{2}(N+2\xi+2\zeta+3)\right)_{m}}
&\text{$N$ odd},
 \end{cases}
\end{align}
and the normalizations are given by
\begin{align}
 \hspace{-2em}
 v_{2m+j}(\xi,\zeta,N)\!=\!
 \begin{cases}
  \!(-1)^{j}2^{4(2m+j)}m!\!\left(\xi\!+\!\tfrac{1}{2}\right)_{m+j}
  \left(\tfrac{1-N}{2}\!-\!\zeta\right)_{m}
  \left(-\tfrac{N}{2}\right)_{m+j}\!
  \frac{\left(-N-\xi-\zeta\right)_{N/2}}{\left(\frac{1-N}{2}-\xi\right)_{N/2}}
  \qquad&\text{$N$ even},\\[1em]
  \!(-1)^{j}2^{4(2m+j)}m!\!\left(\xi\!+\!\tfrac{1}{2}\right)_{m+j}
  \left(\tfrac{1-N}{2}\right)_{m}
  \left(-\zeta\!-\!\frac{N}{2}\right)_{m+j}\!
  \frac{\left(\xi+\zeta+1\right)_{(N+1)/2}}{\left(\zeta+\frac{1}{2}\right)_{(N+1)/2}}
  &\text{$N$ odd}.
 \end{cases}
\end{align}
with $j\in\{0,1\}$ and $m$ an integer.

The dual $-1$ Hahn polynomials are bispectral, but they satisfy a five term difference
relation \cite{Tsujimoto2013} on the grid $x_s$, hence they fall outside the scope of
Leonard duality.

\section{Solutions of the differential equations}\label{sec:differential}

\subsection{The one-dimensional Schr\"odinger equation}\label{subsec:schrodun}
The Schr\"odinger equation of the one-dimensional system is
\begin{align}\label{eq:1dschrod}
 H\psi=\frac{1}{2}\left(-\frac{d^{2}}{d\rho^{2}}+\rho^{2}+\frac{k(k-\sigma_3)}{\rho^{2}}\right)\psi=E\psi.
\end{align}
Let
\begin{align}\label{eq:covstart}
 \psi=e^{-\rho^{2}/2}\rho^{\alpha}f,
\end{align}
where $\alpha$ remains to be fixed.
Putting this back in \eqref{eq:1dschrod} gives
\begin{align}
 \frac{d^{2}f}{d\rho^{2}}+2\left(-\rho+\frac{\alpha}{\rho}\right)\frac{df}{d\rho}+\frac{\alpha(\alpha-1)-k(k-\sigma_3)}{\rho^{2}}f+(2E-2\alpha-1)f=0.
\end{align}
The value of $\alpha$ is now chosen in order to cancel the term in $\rho^{-2}$, that is
\begin{align}
 \alpha=\begin{cases}
         k\phantom{+1}\qquad &\text{if $\sigma_3$ has eigenvalue $s=+1$},\\
         k+1\qquad &\text{if $\sigma_3$ has eigenvalue $s=-1$}.
        \end{cases}
\end{align}
Effecting the change of variable $\rho=x^{1/2}$, this equation becomes
\begin{align}
 x\frac{d^{2}f}{dx^{2}}+\left(\alpha+\frac{1}{2}-x\right)\frac{df}{dx}+\frac{1}{4}\left(2E-2\alpha-1\right)f=0.
\end{align}
The solutions of this equation are identified as generalized Laguerre polynomials \cite{Koekoek2010} $L^{(\beta)}_{n}(x)$ with parameter
\begin{align}
 \beta=\alpha-\tfrac{1}{2}.
\end{align}
The orthonormalized solutions of the one-dimensional system $\psi_{n,k,s}(\rho)$ and the energies $E_n$ are then given by
\begin{align}\label{eq:solut1dlag}
\begin{aligned}
 \psi_{n,k,s}(\rho)=\braket{\rho}{n,k,s}&=\sqrt{\frac{n!}{\Gamma(n+k+1-s/2)}}e^{-\rho^{2}/2}\rho^{k+\frac{1-s}{2}}\,L_{n}^{(k-\frac{s}{2})}(\rho^{2}),\\[0.5em]
 E_n&=2n+k+1-s/2,\qquad s=\pm1.
\end{aligned}
\end{align}
Expressing these solutions in terms of the generalized Hermite polynomials $H_{m}^{k}(x)$ \cite{Chihara1978,Rosenblum1994,Rosler1998}, one obtains \eqref{eq:genHermite}.

\subsection{Separation in polar coordinates}
First start with the angular equation \eqref{eq:polarangulari} and denote
\begin{align}\label{eq:polareigenvalsigma}
 \beta_1=k_1(k_1-s_1),\qquad \beta_2=k_2(k_2-s_2).
\end{align}
This yields
\begin{align}\label{eq:polarangular}
 \left[\frac{d^{2}}{d\phi^{2}}-\frac{\beta_1}{\cos^{2}\phi}-\frac{\beta_2}{\sin^{2}\phi}+m^{2}\right]\Phi=0.
\end{align}
Now take
\begin{align}
 \Phi=\sin^{\gamma}(\phi)\cos^{\delta}(\phi)f,
\end{align}
with $\gamma$ and $\delta$ to be determined, and \eqref{eq:polarangular} becomes
\begin{align}\label{eq:polarangular2}
 \frac{d^{2}f}{d\phi^{2}}+2\left(\gamma\frac{\cos\phi}{\sin\phi}-\delta\frac{\sin\phi}{\cos\phi}\right)\frac{df}{d\phi}
  +\left[\frac{[\gamma(\gamma-1)-\beta_2]}{\sin^{2}\phi}+\frac{[\delta(\delta-1)-\beta_1]}{\cos^{2}\phi}-(\gamma+\delta)^{2}+m^{2}\right]f=0.
\end{align}
The terms in $\cos^{-2}\phi$ and $\sin^{-2}\phi$ are eliminated upon choosing
\begin{align}
 \delta(\delta-1)=\beta_1,\qquad \gamma(\gamma-1)=\beta_2.
\end{align}
Now introduce
\begin{align}
 x=-\cos2\phi,
\end{align}
\eqref{eq:polarangular2} is rewritten as the differential equation
\begin{align}
 (1-x^{2})\frac{d^{2}f}{dx^{2}}+\left[(\gamma-\delta)-(\gamma+\delta+1)x\right]\frac{df}{dx}+\frac{1}{4}\left(m^{2}-(\gamma+\delta)^{2}\right)f=0,
\end{align}
whose solutions are the Jacobi polynomials $P_d^{(\alpha,\beta)}(x)$ with parameters
\begin{align}
 \alpha=\delta-\frac{1}{2},\qquad \beta=\gamma-\frac{1}{2}
\end{align}
and
\begin{align}
 |m|=2d+\gamma+\delta.
\end{align}
Recalling \eqref{eq:polareigenvalsigma}, we finally obtain the orthonormalized angular solution to \eqref{eq:polarangular}
\begin{align}
\begin{aligned}{}
 \Phi_{\ell,s_1,s_2}^{k_1,k_2}(\phi)&=\sqrt{\frac{2(2\ell+k_1+k_2)~(\ell-\frac{2-s_1-s_2}{4})!~\Gamma(\ell+k_1+k_2+\frac{2-s_1-s_2}{4})}{\Gamma(\ell+k_1+\frac{2-s_1+s_2}{4})~\Gamma(\ell+k_2+\frac{2+s_1-s_2}{4})}}\\
 &\qquad\times[\cos\phi]^{k_1+\frac{1-s_1}{2}}[\sin\phi]^{k_2+\frac{1-s_2}{2}}\,P_{\ell-\frac{2-s_1-s_2}{4}}^{(k_1-\frac{s_1}{2},k_2-\frac{s_2}{2})}(-\cos2\phi),\\[.5em]
 |m|&=2\ell+(k_1+k_2),\qquad s_1,s_2\in\{\pm1\}.
\end{aligned}
\end{align}
In the above, $\ell\in\{0,1,2,\dots\}$ is a non-negative integer if $s_1s_2=1$ and $\ell\in\{\frac{1}{2},\frac{3}{2},\dots\}$ is a half-integer if $s_1s_2=-1$.\\[1em]
Next, the radial equation is
\begin{align}
 \frac{d^{2}R}{dr^{2}}+\frac{1}{r}\frac{dR}{dr}+\left(-r^{2}-\frac{m^{2}}{r^{2}}+2E\right)R=0
\end{align}
and its orthonormalized solutions are obtained like the one-dimensional system, see \eqref{eq:covstart}--\eqref{eq:solut1dlag}:
\begin{align}
 R=\sqrt{\frac{2(N!)}{\Gamma(N'+|m|+1)}}r^{|m|}e^{-r^{2}/2}L_{N'}^{(|m|)}(r^{2}).
\end{align}

\newpage
\bibliographystyle{unsrtinurl}
\bibliography{superint_-1Hahn_v3.bib}

\begin{thebibliography}{10}

\bibitem{Genest2013}
V.~X. Genest, L.~Vinet, and A.~Zhedanov.
\newblock {The algebra of dual {$-1$} Hahn polynomials and the Clebsch–Gordan
  problem of {$\mathfrak{sl}_{-1}(2)$}}.
\newblock {\em J. Math. Phys.}, 54(2):1--15, 2013.
\newblock \href {http://arxiv.org/abs/1207.4220} {\path{arXiv:1207.4220}}.

\bibitem{Miller2013}
W.~Miller, S.~Post, and P.~Winternitz.
\newblock {Classical and quantum superintegrability with applications}.
\newblock {\em J. Phys. A Math. Theor.}, 46(42), 2013.
\newblock \href {http://arxiv.org/abs/1309.2694} {\path{arXiv:1309.2694}}.

\bibitem{Letourneau1995}
P.~L{\'{e}}tourneau and L.~Vinet.
\newblock {Superintegrable Systems: Polynomial Algebras and Quasi-Exactly
  Solvable Hamiltonians}.
\newblock {\em Ann. Phys. (N. Y).}, 243(1):144--168, 1995.

\bibitem{Marquette2010}
I.~Marquette.
\newblock {Superintegrability and higher order polynomial algebras}.
\newblock {\em J. Phys. A Math. Theor.}, 43(13), 2010.

\bibitem{DeAlfaro1976}
V.~de~Alfaro, S.~Fubini, and G.~Furlan.
\newblock {Conformal Invariance in Quantum Mechanics}.
\newblock {\em Nuovo Cim.}, 34(4):569--612, 1976.

\bibitem{Genest2014}
V.~X. Genest, L.~Vinet, and A.~Zhedanov.
\newblock {The Racah algebra and superintegrable models}.
\newblock {\em J. Phys. Conf. Ser.}, 512(1):012011, 2014.

\bibitem{Higgs1979}
P.~W. Higgs.
\newblock {Dynamical symmetries in a spherical geometry I}.
\newblock {\em J. Phys. A. Math. Gen.}, 12(3):309, 1979.

\bibitem{Zhedanov1992}
A.~S. Zhedanov.
\newblock {The "Higgs algebra" as a quantum deformation of
  {$\mathfrak{su}(2)$}}.
\newblock {\em Mod. Phys. Lett. A}, 07(06):507--512, 1992.

\bibitem{Bonatsos1995}
D.~Bonatsos, C.~Daskaloyannis, and P.~Kolokotronis.
\newblock {Generalized deformed {$\mathfrak{su}(2)$} algebras, deformed
  parafermionic oscillators and finite W-algebras}.
\newblock {\em Mod. Phys. Lett. A}, 10:2197, 1995.

\bibitem{Granovskii1991}
Y.~I. Granovskii, A.~S. Zhedanov, and I.~M. Lutzenko.
\newblock {Quadratic algebra as a "hidden" symmetry of the Hartmann potential}.
\newblock {\em J. Phys. A. Math. Gen.}, 24(16):3887--3894, 1991.

\bibitem{Granovskii1988}
Y.~A. Granovskii and A.~S. Zhedanov.
\newblock {Nature of the symmetry group of the {$6j$}-symbol}.
\newblock {\em J. Exp. Theor. Phys.}, 94:49--54, 1988.

\bibitem{Frappat2019a}
L.~Frappat, J.~Gaboriaud, L.~Vinet, S.~Vinet, and A.~Zhedanov.
\newblock {The Higgs and Hahn algebras from a Howe duality perspective}.
\newblock {\em Phys. Lett. Sect. A Gen. At. Solid State Phys.},
  383(14):1531--1535, 2019.

\bibitem{Zhedanov1991}
A.~S. Zhedanov.
\newblock {"Hidden symmetry" of Askey–Wilson polynomials}.
\newblock {\em Theor. Math. Phys.}, 89(2):1146--1157, 1991.

\bibitem{Koekoek2010}
R.~Koekoek, P.~A. Lesky, and R.~F. Swarttouw.
\newblock {\em {Hypergeometric Orthogonal Polynomials and Their
  {$q$}-Analogues}}.
\newblock Springer Monographs in Mathematics. Springer Berlin Heidelberg, 2010.

\bibitem{Fubini1984}
S.~Fubini and E.~Rabinovici.
\newblock {Superconformal quantum mechanics}.
\newblock {\em Nucl. Phys. B}, 245:17--44, 1984.

\bibitem{Tsujimoto2013}
S.~Tsujimoto, L.~Vinet, and A.~Zhedanov.
\newblock {Dual {$-1$} Hahn polynomials: “Classical” polynomials beyond the
  Leonard duality}.
\newblock {\em Proc. Am. Math. Soc.}, 141(3):959--970, 2013.

\bibitem{Genest2012}
V.~X. Genest, M.~E. Ismail, L.~Vinet, and A.~Zhedanov.
\newblock {The Dunkl oscillator in the plane: I. Superintegrability, separated
  wavefunctions and overlap coefficients}.
\newblock {\em J. Phys. A Math. Theor.}, 46(14), 2013.
\newblock \href {http://arxiv.org/abs/1212.4459} {\path{arXiv:1212.4459}}.

\bibitem{Genest2014a}
V.~X. Genest, M.~E.~H. Ismail, L.~Vinet, and A.~Zhedanov.
\newblock {The Dunkl Oscillator in the Plane II: Representations of the
  Symmetry Algebra}.
\newblock {\em Commun. Math. Phys.}, 329(3):999--1029, 2014.

\bibitem{Genest2013a}
V.~X. Genest, L.~Vinet, and A.~Zhedanov.
\newblock {The singular and the {$2:1$} anisotropic Dunkl oscillators in the
  plane}.
\newblock {\em J. Phys. A Math. Theor.}, 46(32), 2013.
\newblock \href {http://arxiv.org/abs/1305.2126v2} {\path{arXiv:1305.2126v2}}.

\bibitem{Rosenblum1994}
M.~Rosenblum.
\newblock {Generalized Hermite Polynomials and the Bose-Like Oscillator
  Calculus}.
\newblock {\em Oper. Theory Adv. Appl.}, 73:369--396, 1994.

\bibitem{Mukunda1980}
N.~Mukunda, E.~C.~G. Sudarshan, J.~K. Sharma, and C.~L. Mehta.
\newblock {Representations and properties of para-Bose oscillator operators. I.
  Energy position and momentum eigenstates}.
\newblock {\em J. Math. Phys.}, 21(9):2386--2394, 1980.

\bibitem{Dunkl1991}
C.~F. Dunkl.
\newblock {Integral kernels with reflection group invariance}.
\newblock {\em Can. J. Math.}, 43(6):1213--1227, 1991.

\bibitem{Rodriguez2008}
M.~A. Rodr{\'{i}}guez, P.~Tempesta, and P.~Winternitz.
\newblock {Reduction of superintegrable systems: The anisotropic harmonic
  oscillator}.
\newblock {\em Phys. Rev. E}, 78(4):046608, 2008.

\bibitem{Bergeron2016}
G.~Bergeron and L.~Vinet.
\newblock {Generating functions for the {$\mathfrak{osp}(1|2)$}
  Clebsch–Gordan coefficients}.
\newblock {\em J. Phys. A Math. Theor.}, 49(11):115202, 2016.
\newblock \href {http://arxiv.org/abs/1507.00018} {\path{arXiv:1507.00018}}.

\bibitem{Genest2015a}
V.~X. Genest, L.~Vinet, and A.~Zhedanov.
\newblock {A Laplace-Dunkl Equation on {$S^2$} and the Bannai–Ito Algebra}.
\newblock {\em Commun. Math. Phys.}, 336(1):243--259, 2015.

\bibitem{Tsujimoto2011}
S.~Tsujimoto, L.~Vinet, and A.~Zhedanov.
\newblock {From {$sl_q(2)$} to a Parabosonic Hopf Algebra}.
\newblock {\em Symmetry, Integr. Geom. Methods Appl.}, 7(093):13, 2011.
\newblock \href {http://arxiv.org/abs/1108.1603v3} {\path{arXiv:1108.1603v3}}.

\bibitem{Frappat1996}
L.~Frappat, A.~Sciarrino, and P.~Sorba.
\newblock {\em {Dictionary on Lie Superalgebras}}.
\newblock Academic Press, 1996.
\newblock \href {http://arxiv.org/abs/hep-th/9607161}
  {\path{arXiv:hep-th/9607161}}.

\bibitem{Chihara1978}
T.~S. Chihara.
\newblock {\em {An Introduction to Orthogonal Polynomials}}.
\newblock Gordon and Breach, New-York, 1978.

\bibitem{Rosler1998}
M.~R{\"{o}}sler.
\newblock {Generalized Hermite polynomials and the heat equation for Dunkl
  operators}.
\newblock {\em Commun. Math. Phys.}, 192(3):519--542, 1998.

\bibitem{Rainville1960}
E.~D. Rainville.
\newblock {\em {Special functions}}.
\newblock The Macmillan Company, New York, 1960.

\bibitem{Schluter1983}
P.~Schluter, K.~H. Wietschorke, and W.~Greiner.
\newblock {The Dirac equation in orthogonal coordinate systems. I. The local
  representation}.
\newblock {\em J. Phys. A. Math. Gen.}, 16(9):1999--2016, jun 1983.

\bibitem{Gaboriaud2018b}
J.~Gaboriaud, L.~Vinet, S.~Vinet, and A.~Zhedanov.
\newblock {The dual pair {$Pin(2n)\otimes\mathfrak{osp}(1|2)$}, the Dirac
  equation and the Bannai–Ito algebra}.
\newblock {\em Nucl. Phys. B}, 937:226--239, 2018.
\newblock \href {http://arxiv.org/abs/1810.00130} {\path{arXiv:1810.00130}}.

\bibitem{Kalnins1996}
E.~G. Kalnins, W.~Miller, and G.~S. Pogosyan.
\newblock {Superintegrability and associated polynomial solutions: Euclidean
  space and the sphere in two dimensions}.
\newblock {\em J. Math. Phys.}, 37(12):6439--6467, 1996.

\bibitem{Kalnins2007}
E.~G. Kalnins, W.~Miller, and S.~Post.
\newblock {Wilson polynomials and the generic superintegrable system on the
  {$2$}-sphere}.
\newblock {\em J. Phys. A Math. Theor.}, 40(38):11525--11538, 2007.

\bibitem{Kalnins2013}
E.~G. Kalnins, W.~Miller, and S.~Post.
\newblock {Contractions of {$2D$} 2nd Order Quantum Superintegrable Systems and
  the Askey Scheme for Hypergeometric Orthogonal Polynomials}.
\newblock {\em Symmetry, Integr. Geom. Methods Appl.}, 9:57--84, 2013.

\bibitem{Genest2014f}
V.~X. Genest, L.~Vinet, and A.~Zhedanov.
\newblock {The Bannai–Ito algebra and a superintegrable system with
  reflections on the two-sphere}.
\newblock {\em J. Phys. A Math. Theor.}, 47(20):205202, 2014.

\bibitem{Genest2013d}
V.~X. Genest, L.~Vinet, and A.~Zhedanov.
\newblock {Bispectrality of the Complementary Bannai–Ito polynomials}.
\newblock {\em Symmetry, Integr. Geom. Methods Appl.}, 9:18, 2013.

\bibitem{Lemay2018}
J.-M. Lemay and L.~Vinet.
\newblock {Bivariate Bannai–Ito polynomials}.
\newblock {\em J. Math. Phys.}, 59(12), 2018.

\bibitem{DeBie2016b}
H.~{De Bie}, V.~X. Genest, J.-M. Lemay, and L.~Vinet.
\newblock {A superintegrable model with reflections on {$S^3$} and the rank two
  Bannai–Ito algebra}.
\newblock {\em Acta Polytech.}, 56(3):166--172, 2016.

\bibitem{Harnad2004}
J.~Harnad and O.~Yermolayeva.
\newblock {Superintegrability, Lax matrices and separation of variables}.
\newblock In {\em CRM Proc. Lect. Notes}, volume~37, chapter~6, pages 65--73.
  2004.
\newblock \href {http://arxiv.org/abs/nlin/0303009}
  {\path{arXiv:nlin/0303009}}.

\bibitem{Crampe2019a}
N.~Cramp{\'{e}}, L.~Vinet, and M.~Zaimi.
\newblock {Bannai–Ito algebras and the universal {$R$}-matrix of
  {$\mathfrak{osp}(1|2)$}}.
\newblock pages 1--9, 2019.

\end{thebibliography}

\end{document}